\documentclass[useAMS,usenatbib,usegraphicx]{mn2e}

\title[ULTRACAM observations of FS Aurigae]
{Steps towards a solution of the FS Aurigae puzzle.\\
I. Multicolour high-speed photometry with ULTRACAM.}
\author[V.\,V.\,Neustroev et al.]{V.\,V.\,Neustroev$^{1}$\thanks{E-mail:
benj@it.nuigalway.ie},
S.\,Zharikov$^2$, G.\,Tovmassian$^2$\thanks{Visiting
research fellow at Center for Astrophysics and Space Sciences (CASS),
University of California, San Diego (UCSD), 9500 Gilman Drive,
La Jolla, CA 92093-0424, USA} and A.\,Shearer$^1$\\
$^{1}$Computational Astrophysics Laboratory, National University
of Ireland, Galway, Newcastle Rd., Galway, Ireland\\
$^{2}$Observatorio Astronomico Nacional, Instituto de Astronomia, UNAM,
P.O. Box 439027, San Diego, CA 92143-9027}
\begin{document}

\date{Accepted ???. Received ???; in original form 2005 January 31}

\pagerange{\pageref{firstpage}--\pageref{lastpage}} \pubyear{2005}

\maketitle

\label{firstpage}

\begin{abstract}

We present the analysis of high-speed photometric observations
of FS\,Aur taken with the 4.2-m William Herschel Telescope and the high-speed
camera ULTRACAM in late 2003.
These observations were intended to determine whether
there was any evidence of photometric variability with a period in the range
50--100 seconds. The discovery of such variations would help to explain existence,
in FS Aur, of the very coherent photometric period of 205.5 min that exceeds the
spectroscopic period by 2.4 times. Such a discrepancy in the photometric and
spectroscopic periods is an unusual  for a low mass
binary system that is unambiguously identified as a Cataclysmic Variable.
Using various methods, including wavelet-analysis, we found that
with exception of the 205.5-minute periodicity, the main characteristic
of variability of FS Aur is usual flickering and Quasi-Periodic Oscillations.
However, we  detected variability with a  period of $\sim$101 and/or $\sim$105 sec,
seen for a short time every half of the orbital period.
These oscillations may be associated  with the spin period
of the white dwarf, not ruling out the possibility that we are observing a precessing
rapidly rotating white dwarf in FS Aur.

\end{abstract}


\section{Introduction}

Cataclysmic Variables (CVs) are close interacting binaries that contain a white
dwarf accreting material transferred from a companion, usually a late main-sequence
star. CVs are very active photometrically, exhibiting variability on time scales
from seconds to centuries (see review by \citealt{warner}).
In this paper we concentrate on FS Aurigae, a cataclysmic
variable with rather unusual photometric behaviour.

FS Aur was discovered and first classified as a dwarf nova by \citet{hoff}.
This system varies between approximately V=15$\fm$4--16$\fm$2 in
quiescence and V=14$\fm$4 in outburst.
\citet{TPST} defined the orbital period of 0.0595 days (=85.7 min) from H$\alpha$
spectroscopy and suggested an SU UMa classification based on the short orbital
period:
``In view of the preponderance of SU UMa stars among the dwarf novae in this
period range, we would be surprised if superoutbursts and superhumps were not
detected in the future''.
However no superoutburst/superhumps have ever been observed in FS~Aur.
So the SU Uma classification of this star is uncertain.

\begin{table*}
 \centering
 \begin{minipage}{150mm}
  \renewcommand{\thefootnote}{\alph{footnote}}
  \caption{Log of ULTRACAM observations.}
  \begin{tabular}{@{}cclcccl@{}}
  \hline
   Date     &
UT\footnotemark[1]        &
Run   & Number\footnotemark[1] & Exposure & Filters &   Comments\\
            &   start - end    &         & of points      &  (sec)   &         &           \\
\hline
2003-10-30 & 00:57:39 - 01:38:00 &   run021(a) &   484  & 5.01  &    ugi   &     Cirrus and poor seeing ($>$3\arcsec)   \\
2003-10-30 & 01:58:28 - 02:36:12 &   run021(b) &   450  & 5.01  &    ugi   &     Cirrus and seeing improving            \\
2003-10-30 & 02:37:36 - 03:19:45 &   run023    &   506  & 5.01  &    ugi   &                                            \\
2003-10-30 & 04:06:16 - 04:20:38 &   run026    &   157  & 5.01  &    ugi   &     Seeing around 1\arcsec. Hit rotator limit    \\
2003-10-30 & 04:36:58 - 05:39:13 &   run027(a) &    28  & 5.01  &    ugi   &     Cirrus again                           \\
2003-10-30 & 05:45:09 - 05:49:27 &   run027(b) &   727  & 5.01  &    ugi   &     Cirrus and seeing improving            \\
\hline
2003-11-03 & 03:10:50 - 03:40:45 &   run065    &   824  & 2.18  &    ugr   &                                            \\
2003-11-03 & 03:40:57 - 04:10:07 &   run066    &   803  & 2.18  &    ugr   &     Run stopped for zenith blind-spot slew \\
2003-11-03 & 04:13:35 - 04:44:08 &   run067    &   841  & 2.18  &    ugr   &                                            \\
2003-11-03 & 04:44:17 - 05:14:21 &   run068    &   828  & 2.18  &    ugr   &                                            \\
2003-11-03 & 05:14:29 - 05:44:14 &   run069    &   819  & 2.18  &    ugr   &                                            \\
2003-11-03 & 05:44:25 - 06:14:34 &   run070    &   830  & 2.18  &    ugr   &                                            \\
2003-11-03 & 06:15:02 - 06:16:17 &   run071    &   35   & 2.18  &    ugr   &     Dome closed due to high humidity       \\
\hline
\end{tabular}

\footnotetext[1]{Due to cirrus and poor seeing during first night (29/30 October)
some sets of data were not useful for an analysis.
The times of observations (UT) and the ``number of points'' presented in this table
are the actual values used for the following studying.}
\end{minipage}
\label{Table}
\end{table*}

Moreover, recently another more outlandish peculiarity of FS Aur has been
discovered. \citet{Neustroev} in the first detailed investigation of this
system noted,
and \citet{T03} confirmed on the base of a large dataset spanning for several
years that this system shows a well-defined {\it photometric} optical modulations
with 205.5 minutes periodicity and the amplitude of 0.24 mag in B, V and R bands.
This is in contrast to the {\it spectroscopic} period of 85.7 minutes, also confirmed
by \citet{Neustroev} and \citet{T03}.

In addition to FS~Aur, there are a few other CVs with a photometric period that exceeds
the spectroscopic one. These are GW~Lib \citep{Woudt-Warner}, Aqr1 \citep{WWP} and probably
SDSS 1238 \citep{Zharikov}. There is no obvious relationship between photometric
and spectroscopic periods of these objects. However, due to insufficient
observational coverage of the latter objects it is unclear how stable and coherent the
photometric modulations are. For example, the 125.4-min modulation in
the best known system GW~Lib were observed for only two weeks and were not detected
in other observations.
At the same time, the photometric modulations of FS Aur have been observed to be
coherent from ten years of observations  (\citealt{T04}; Neustroev et al., in preparation).
%It states FS Aur as the unique system with not explained while behaviour.
On the other hand several high frequency periodic signals were detected
in the light curve of GW Lib \citep{GWLib}  and it has been argued that they are
ZZ Cet type pulsations of a WD. More observations are needed to determine the similarities
and differences of these objects.

The reason for the strange behaviour of FS~Aur is not yet clear.
\citet{T03}  concluded that the
``usual'' effects explaining photometric period that are longer than
spectroscopic, can be effectively ruled out.
In particular, it is difficult to directly apply the intermediate polar
scenario to the case of FS Aur. There are theoretical restrictions on upper
limits of the spin periods of IPs, and 205 min is very unlikely to be
the period of rotation of the WD.

\citet{T03} also showed that a possible explanation of this puzzle might be
a rapidly rotating magnetic white dwarf in FS Aur freely precessing with
a period equal to the observed \textit{photometric} period.
The chance to see this precession period in optical wavelengths would only
be when the collimated beam from the magnetically accreting white
dwarf, the rotational axes of the white dwarf and the binary orbital plane
would have certain angles relative to each other. This could explain the
uniqueness of FS Aur and why we observe such periods so rarely.

In order to have the proposed precession period, the rotational white dwarf spin
period should be of the order of 50--100 sec, according to existing models
\citep{Lei92}, i.e. the white dwarf in FS\,Aur could be a fast rotator such as
AE\,Aqr and DQ\,Her. This hypothesis remains, however, highly debatable
until firm evidence of the fast rotating white dwarf in this system is found.
We suggest that this phenomenological model can be observationally
tested with fast optical photometry. Indeed, if we accept that the
photometric variability comes from the free precession of the magnetic white
dwarf, or rather reflection of the high energy beam originating at accreting
poles from elsewhere in the system, then one can reasonably expect to
see variability associated with the rotation period itself.
Furthermore we might expect a relationship between the precessional,
orbital and rotational periodicity.

In this paper we present the analysis of high-speed photometric observations
of FS\,Aur which were intended to determine whether such fluctuations
exist, and we report new interesting features of this system.

\begin{figure}
\includegraphics[width=84mm]{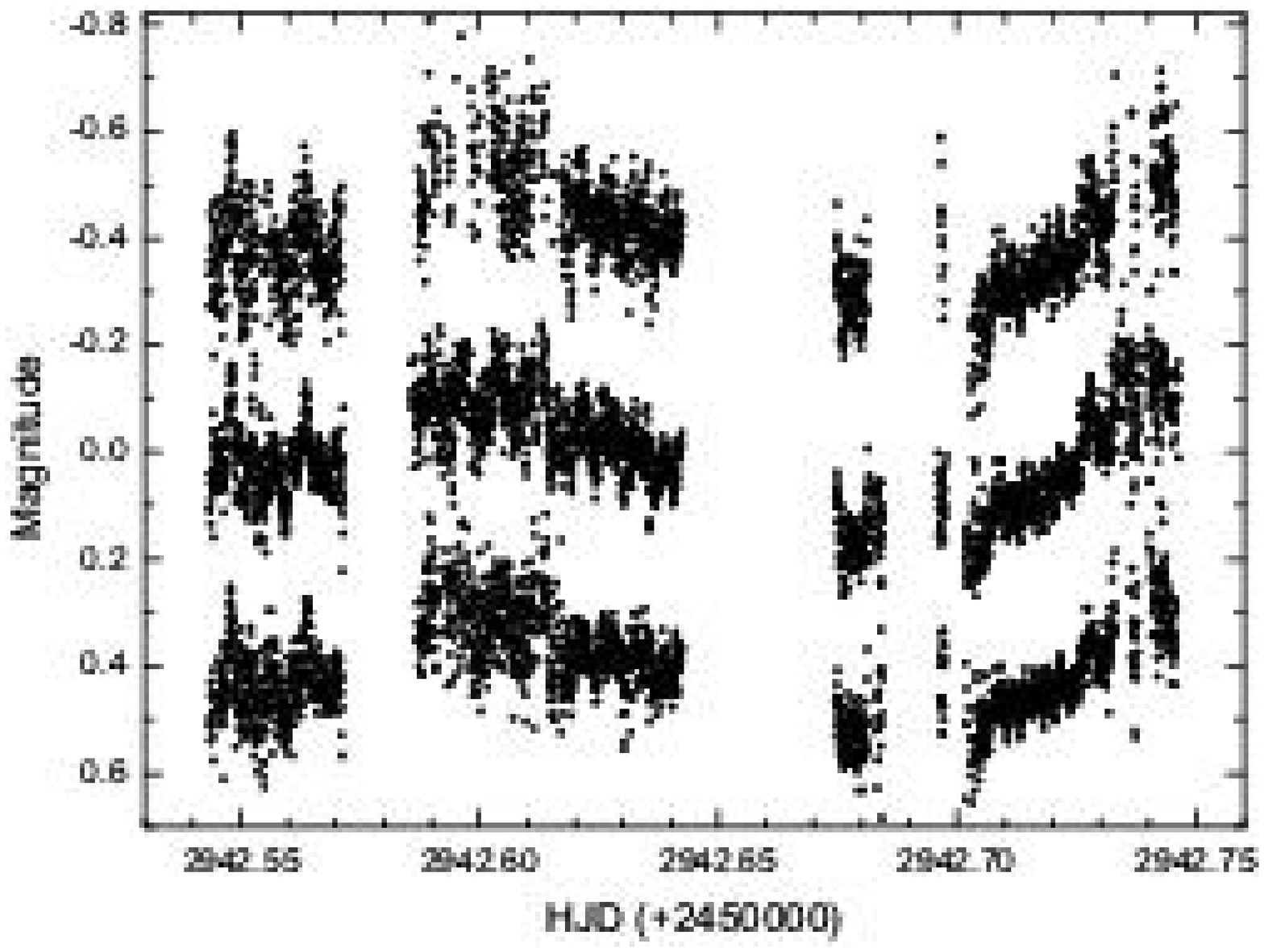} \\
\includegraphics[width=84mm]{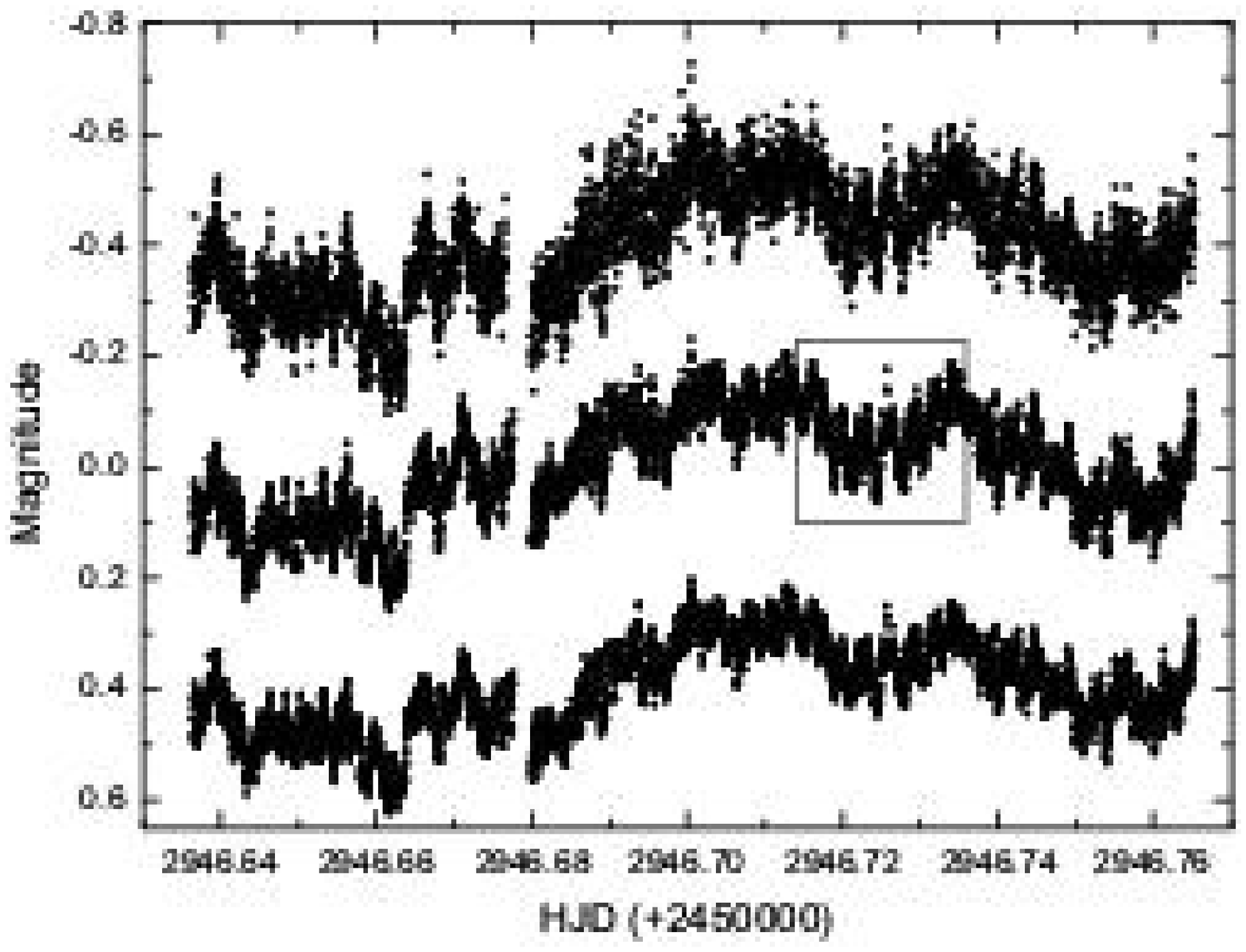} \\
\includegraphics[width=84mm]{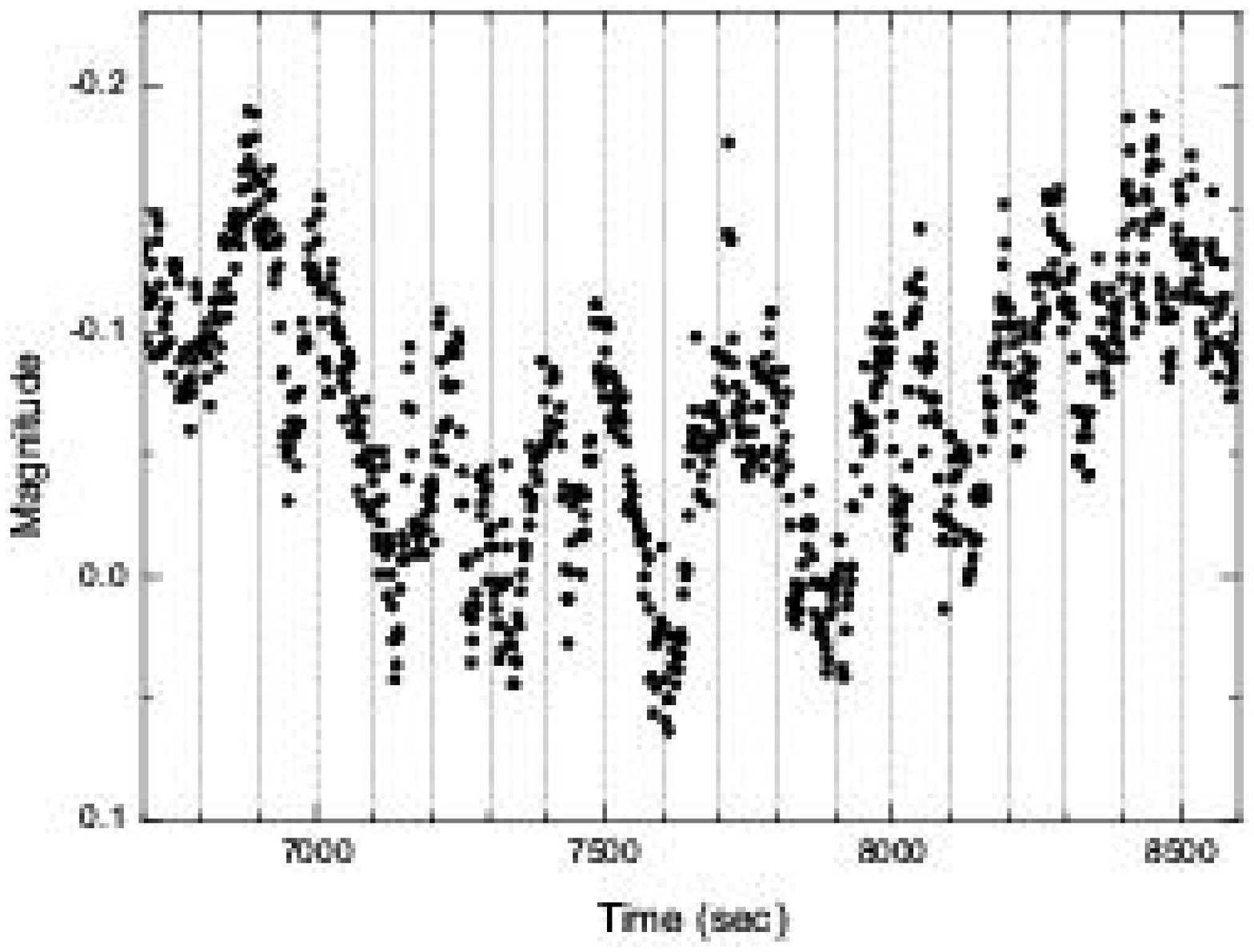}
\caption{The ULTRACAM light curves of FS Aur, obtained on the nights
of 29 October 2003 (upper panel) and 2 November 2003 (middle panel).
The data for the each colour band were normalized to its average
nightly magnitude, after that a vertical shift of 0.4 mag has been introduced
between light curves for display purposes. Upper and medium light curves
are for the $u'$ and $g'$ filters, respectively, and lower curves are
for the $i'$ filter (first night, upper panel) and $r'$ filter (second
night, middle panel).
% Dashed lines denote borders of the different
% subsets (see text).
Bottom panel shows the marked part of the $g'$-band
light curve from second night on a larger scale.}
\label{LightCurves}
\end{figure}

\section{Observations}

FS Aur was observed on the nights of 29 October 2003 and
2 November 2003 using ULTRACAM in Service mode on the 4.2-m
William Herschel Telescope (WHT) at the Isaac Newton
Group of Telescopes, La Palma. ULTRACAM is an ultra-fast,
triple-beam CCD camera designed to provide imaging photometry
at high temporal resolution in three different colours simultaneously
(for further details see \citealt{DhillonMarsh}).
Our observations were obtained
simultaneously in the SDSS $u'$, $g'$, and $i'$ colour bands on
29 October (hereinafter we call this set of observations
as \textit{first night}), and the SDSS $u'$, $g'$, and $r'$
bands on 2 November (\textit{second night}).
The $u'$, $g'$, $r'$ and $i'$ filters have effective wavelengths
of 3550 \AA , 4750 \AA, 6200 \AA\, and 7650 \AA\, respectively.

The weather conditions during the first night were
not optimal due to cirrus and poor seeing (down to 3\arcsec).  We removed the poor
data from the first dataset, and used the rest for this
investigation. During the second night the weather conditions
were on the whole good. This, almost evenly sampled dataset,
of more than three hours duration has very small gaps.
For the first night, we used an exposure time of 5.0 sec;
on the second night the exposure time was decreased to 2.18 sec.

A log of the observations is given in Table~\ref{Table}.

Data reduction was carried out using the \verb"MIDAS" package in
conjunction with Fortran code\footnote{
This Fortran code, developed by Marten van Kerkwijk, can be used
for extracting frames from the ULTRACAM data and converting then
to the MIDAS files.}.
All extracted images were debiased and
flat-fielded in a standard fashion within \verb"MIDAS".
Unfortunately, we were unable to observe a standard star using ULTRACAM,
so the zeropoints for all the filters could not be calculated.
However, this will only affect the scale of the light curves,
not their shape. The usual and previously used comparison stars
C1, C3 and C4 from \citet{Misselt} were included in the field
of view so that differential photometry could be obtained.
We calculated magnitudes of the source with respect to the C1,
using aperture photometry.
As a check of the photometry and systematics in the reduction and
extraction procedures, we also computed magnitudes of the comparison
stars C3 and C4.
For the second night the differences between nightly averages of
the differential magnitudes (C1-C3, C1-C4) were within 0.03 mag
for the $g'$ and $r'$ filters and 0.1 for the $u'$ filter.
The same differences for the used data of the first night
were within 0.05 mag for the $g'$ and $i'$ filters and 0.1
for the $u'$ filter.
The appropriate standard deviations of these differences were
within 0.015 mag for $g'$, $r'$ and $i'$ and 0.05 for $u'$ for
both observational nights.

In this paper we also used data from our previous observations taken with the 1.5-m telescope of OAN SPM  on JD 2451946, we will refer to this
dataset as \textit{set-1946}. Though this data has a time resolution worse
than the ULTRACAM-data ($\sim$20~sec), it covers a longer time span
(315 minutes). All details on these observations can be found in \citet{T03}.

\section{ULTRACAM light curve analysis}
\subsection{Light Curve Morphology}
Fig.~\ref{LightCurves} shows the ULTRACAM light curves of FS Aur.
The most prominent feature is the quasi-sinusoidal shape.
These light curves can be easily
combined with older photometrical data of FS Aur,
confirming the very coherent nature of the periodic signal with the period of $\sim$205.5 minutes
covering ten years of observations (Neustroev et al., in preparation).
Additionally, both datasets show strong modulations
(with an amplitude up to 0.15-0.20 mag) on both medium
($\sim$10 min) and short ($\sim$100 sec) timescales.
This variability is evident in all filters. Moreover,
all the light curves are very correlated and the modulations
in each have almost the same amplitude.

On both nights the mean V magnitude was about 15.7, calculated
using the transformation formulae between the UBVRI and SDSS systems
of \citealt{Smith} indicating that FS Aur was definitely in the quiescent state.

In the bottom panel of Fig.~\ref{LightCurves} we show part of the $g'$-band light curve from
second night. A period of the order of $\sim$100~s is quite apparent to
the eye, but on closer inspection the variability appears to be very complex - the interval between maxima is not constant.

In the following analysis, due to the better quality and more homogeneous nature
of the data, we shall use data from the second night and then compare these results with the first night.

\subsection{Power Spectra}

\begin{figure}
\includegraphics[width=84mm]{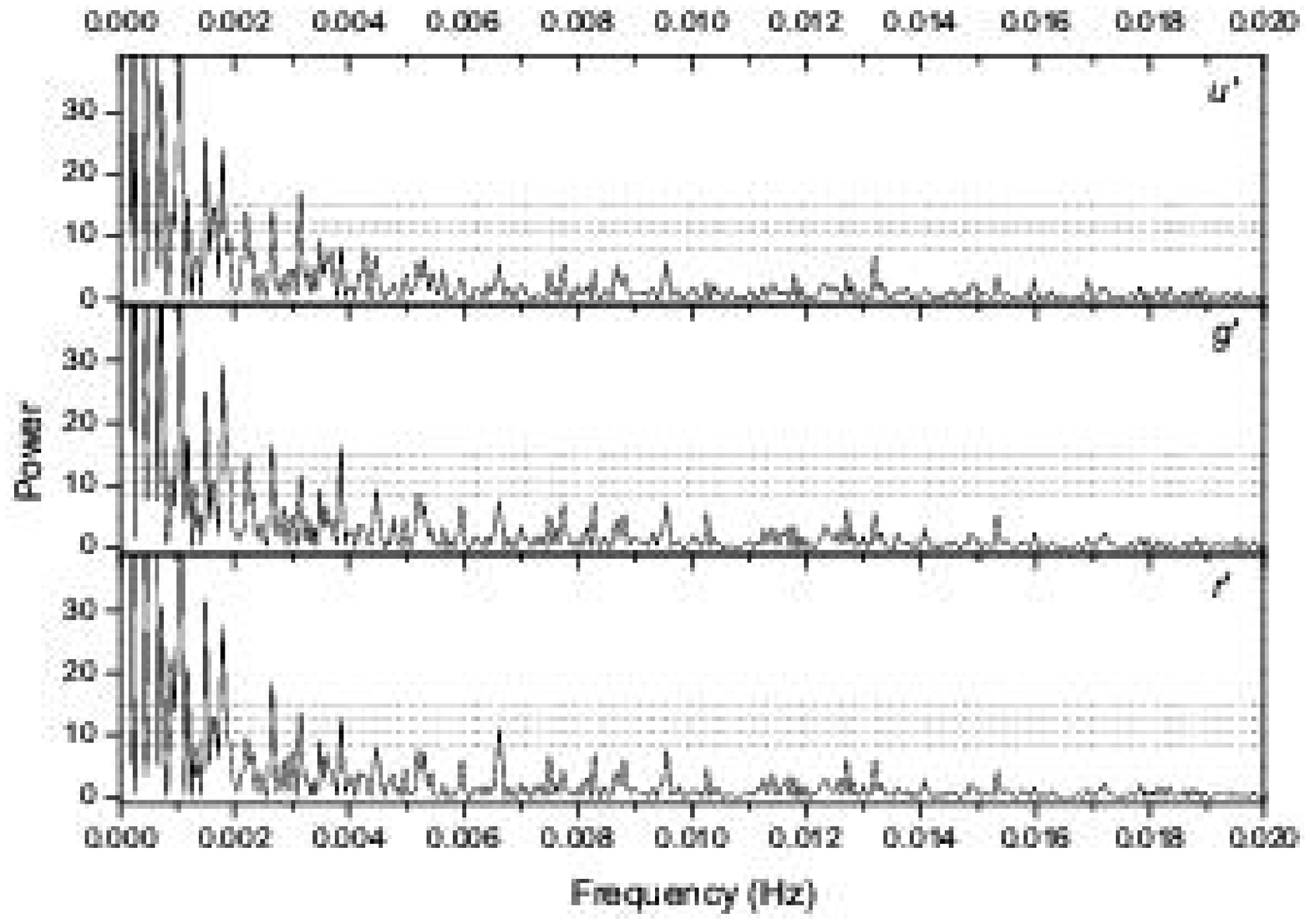}
\vspace{5mm}
\includegraphics[width=84mm]{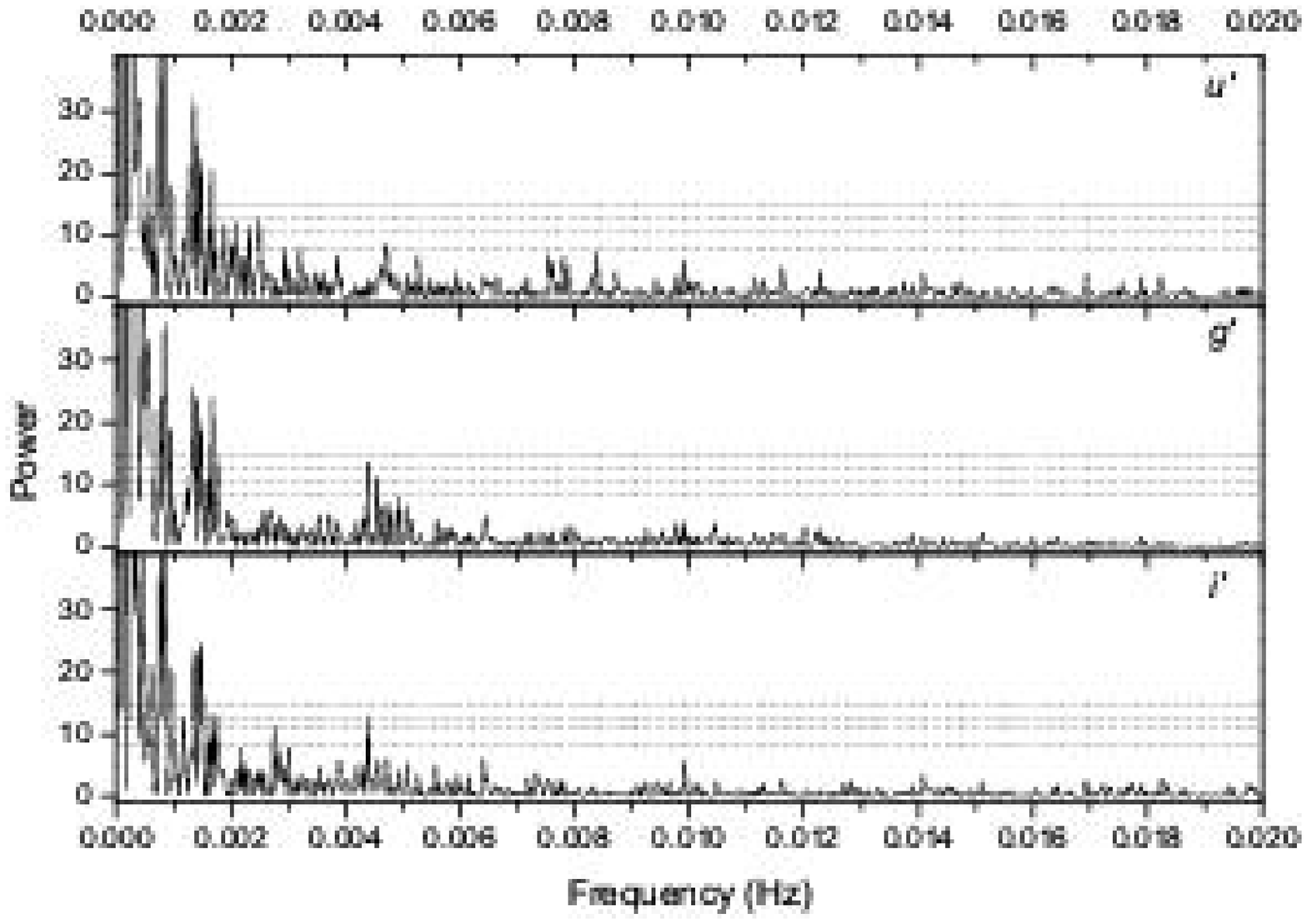}
\caption{The power spectra of the second (upper) and first (bottom) night's light curves.
The overplotted dashed lines are 50\%, 95\%, 99\% and 99.9\%
significance levels.}
\label{power}
\end{figure}

Using the standard Lomb-Scargle power spectrum analysis \citep{Lomb, Scargle},
we calculated the power spectra on individual nights (Fig.~\ref{power},
note the high coincidence between the power spectra in the different filters).
There are a number of strong peaks in the spectra.
In order to understand the significance of these peaks are, we evaluated
their  significance level using a classical method based on white noise
\citep{NumRec}. However we note that the presence of red-noise will alter the significance
levels compared to Gaussian noise models, see discussion in Section~\ref{red-noise}.

All the power spectra are dominated by the fundamental
frequency of $1/P_{phot}$ ($8.1\times10^{-5}$ Hz) and its harmonics.
However there also are the other strong peaks with a  high significance level.
But after comparing, one can see that the power spectra from the different nights
are significantly different. Almost all the strongest peaks
of one night are completely absent in the other and vice versa,
and only few from them have weak counterparts ($\sim$1.50~mHz and 3.86~mHz).
In the frequency range of interest ($\sim$10--20 mHz) one can see,
in both night's power spectra, some relatively strong peaks around 10 mHz.
However, their frequencies do not coincide exactly.

In order to investigate a transition and a mutation of the peaks,
we split up each light curve into four sets of almost equal length
and analyzed all independent sets individually.
None of the calculated spectra is similar to any other.
The above mentioned peaks appear/disappear very quickly,
on the subsequent data sets, so their transition time is very short.
The reason for this is probably simple flickering seen in many CVs.

\subsection{Sliding periodograms and Wavelet analysis}

In order to study how stable the peaks in the power spectra are, the common
practice is to split  the complete time series into a few separate sets and analyse them individually.
Frequently unjustified significance is given to the lack of some
peaks in some subsets with these peaks being eliminated from further analysis.
However it may be important to retrace changes
of these peaks with time. For this we can calculate power spectra for
complete time-series as well as power spectra from a \textit{sliding window}
moved across the full time series with a width of $\Delta$T.

Such an approach can be described, using a
\textit{short time Fourier transform} or \textit{windowed Fourier transform}, \citet{Gabor}:

\begin{equation}
\label{Gabor}
GT(\nu,t_0,\sigma) = \int\limits_{- \infty}^\infty{f(t)e^{-\frac{{(t-t_0)^2}}{{\sigma^2}}}e^{-i2\pi \nu t}dt}
\end{equation}

\noindent
where constant $\sigma$ determines the effective width of the Gabor
transform window, and $t_0$ is the centre of the latter. Moving the
centre of the window along the time series $f(t)$, allows us to obtain
``snapshots'' of the time-frequency behaviour of $f(t)$.
The effective width $\sigma$ determines length of an interval $\Delta$T
which gives the main contribution to value of integral~(\ref{Gabor}).
Here it is important to note that $\Delta$T is a measure of the time
resolution while the linewidth $\Delta \nu$ determines a measure of
the frequency resolution.

Since $\Delta \nu$ and $\Delta$T are inversely related, there is
a problem of the choice of window width. Too
broad a window can provide a reasonable representation of low-frequency
components of the time series, but its width will be redundant for high
frequency harmonics, as all irregularities in high-frequency area of a
spectrum will smooth out. On the contrary, a narrow window will enable us to
study variations in time of the high-frequency features, but it will not
be adequate for low-frequency harmonics. However, having long enough time
series with high time resolution a compromise can be reached.

On the other side, wavelet analysis \citep{Mallat} offers an alternative to Fourier
based time-series analysis and it is particularly useful when spectral
components are time dependent. The wavelet transform represents
one-dimensional signals as a function of both time and frequency and
is similar in this regard to a windowed Fourier transform, but, with one major
difference, namely the window function depends on frequency so
that for low frequencies the window became wider, and for high
frequencies - narrower. Consequently the wavelet transform effectively allows for a multi-scale analysis
of a time series.

Standard wavelet methods require time series to be evenly spaced.
\citet{Foster} developed the weighted wavelet Z-transform (WWZ) that
appears to be an effective method of the analysis of unevenly sampled data.
The idea behind this technique is basically the same as that behind the
Lomb-Scargle periodogram: wavelet trial functions are extended and
corrected by a rescale procedure to satisfy the admissibility condition on
the uneven grid of times of a data record. In this case the wavelet
transform becomes a projection onto trial functions that
resemble the shape of the data record (see also an analysis of this method by
\citealt{Haubold}).

In spite of the fact that the WWZ method seems to have an advantage over a
Fourier transform in our analysis we start with the latter and then go onto the Wavelet anaysis. There are two
reasons for this. Firstly, many nuances of the WWZ are unknown.
For example, the decay constant $c$ in this
method defines the width of the wavelet ``window''. Smaller
values of $c$ will produce wider windows. Using small values of $c$ will
result in improved frequency resolution of variations, but will smear out
temporal variations. Conversely, large values of $c$ will improve the
temporal resolution, but will generate larger uncertainties in peak
frequency. So criteria for choosing  $c$ are not quite clear.
Secondly, we need a direct comparison of  these new calculations with
previously obtained Lomb-Scargle periodograms.

\subsubsection{Sliding periodograms}

To obtain the sliding periodogram (i.e. the full set of the power
spectra calculated with the window moving along all the time series) we calculate
the Lomb-Scargle periodogram in the selected
window moving along the time series, rather than compute integral (\ref{Gabor}).
Formally for the case of evenly spaced data $f(t_i), i=1,...,N$
it can be described as

\begin{eqnarray}
\label{SP}
SP(\nu,T_i,N_W) = LSP(\nu,i,i+N_W-1) \\
i=1 \rightarrow N-N_W+1 \nonumber
\end{eqnarray}

\noindent
where $N_W$ is number of the points in the window,
$LSP$ is the Lomb-Scargle periodogram, calculated in the window
formed from a data point $i$ up to a data point $i+N_W-1$, and
$T_i$ is the effective time of the window centre

\begin{equation}
T_i  \equiv \frac{1}{{N_W}}\sum\limits_{j=i}^{i+N_W-1}{f(t_j)},
\qquad i=1 \rightarrow N-N_W+1
\end{equation}

\noindent

If necessary equation (\ref{SP}) can be easily extended for
the case of unevenly sampled data.
%, especially for long time series with high time resolution.

This approach has obvious advantages over computing the
integral (\ref{Gabor}). Namely, calculated periodograms in the
every window are the usual Lomb-Scargle periodograms and can
be directly compared with periodograms obtained on the complete
sequence of data. We can also easily determine the
reliability of any features in a periodogram. Such approach is
much faster.

Before calculating the sliding periodogram for the second
night we have preprocessed the data.
In order to obtain nearly evenly sampled values of $T_i$, due
to some small gaps in the time series, these were filled by
the mean value of the data.
We have also experimented with different numbers of
points $N_W$ in the window, which defines the tradeoff between
time resolution and frequency resolution, and settled on $N_W$=500 as
a good compromise. Such window has a time span of about 1000 sec.
And finally, the time series was detrended by subtraction of the smoothed
dataset formed from the original by averaging 300 adjacent
data points. This was valid as the periodogram calculated
for the detrended dataset, shows the same features with almost the same
strength. The exceptions were the components of lowest frequency,
which could not be investigated by the proposed method.

We have also obtained a sliding periodogram for the first night.
Since these data have several large gaps we have split the data into
three continuous sets and analyzed them separately, then combined all
the results into one the sliding periodogram. Moreover, in order to
obtain the same frequency resolution we decreased the numbers of
points in the window $N_W$ to 200.

\begin{figure*}
\vspace{5mm}
\includegraphics[width=99mm]{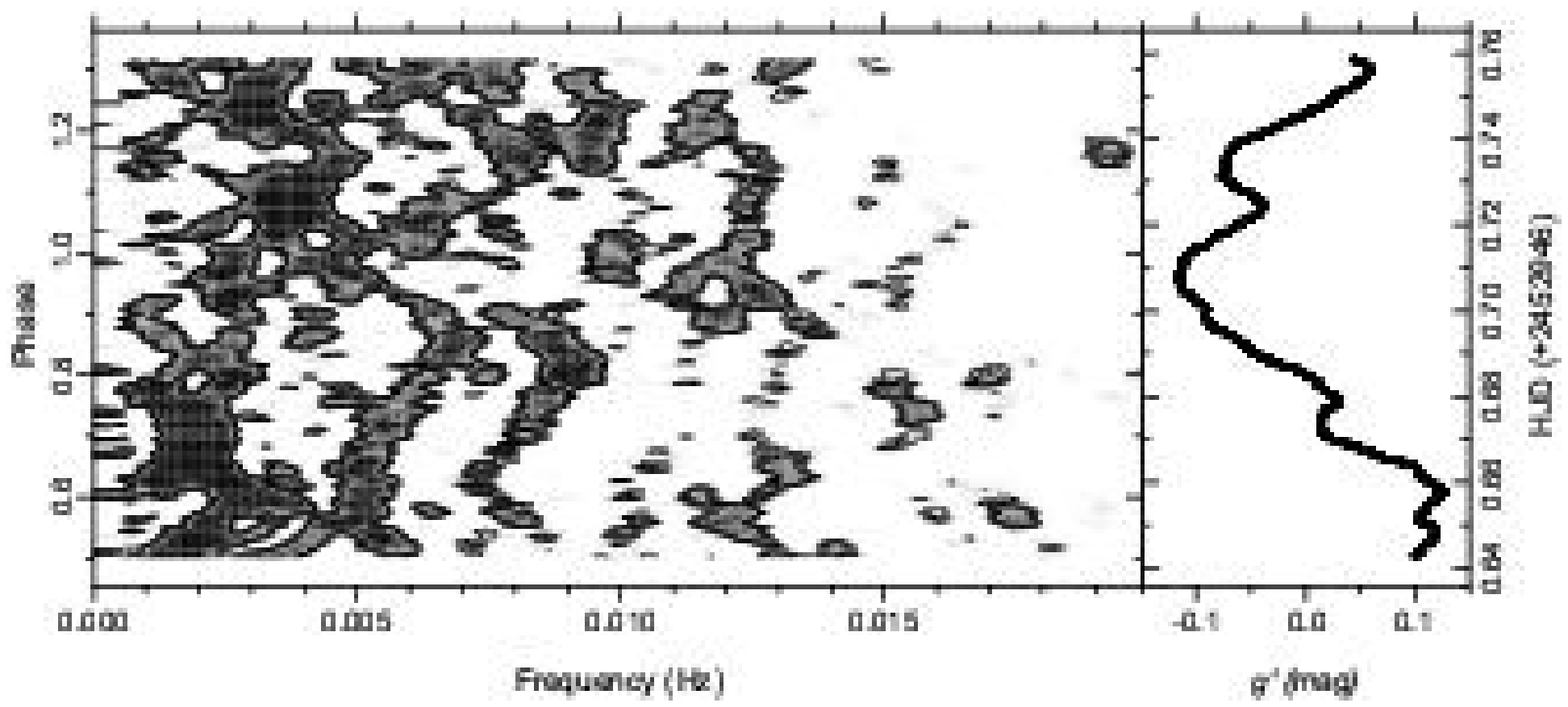}
\includegraphics[width=75mm]{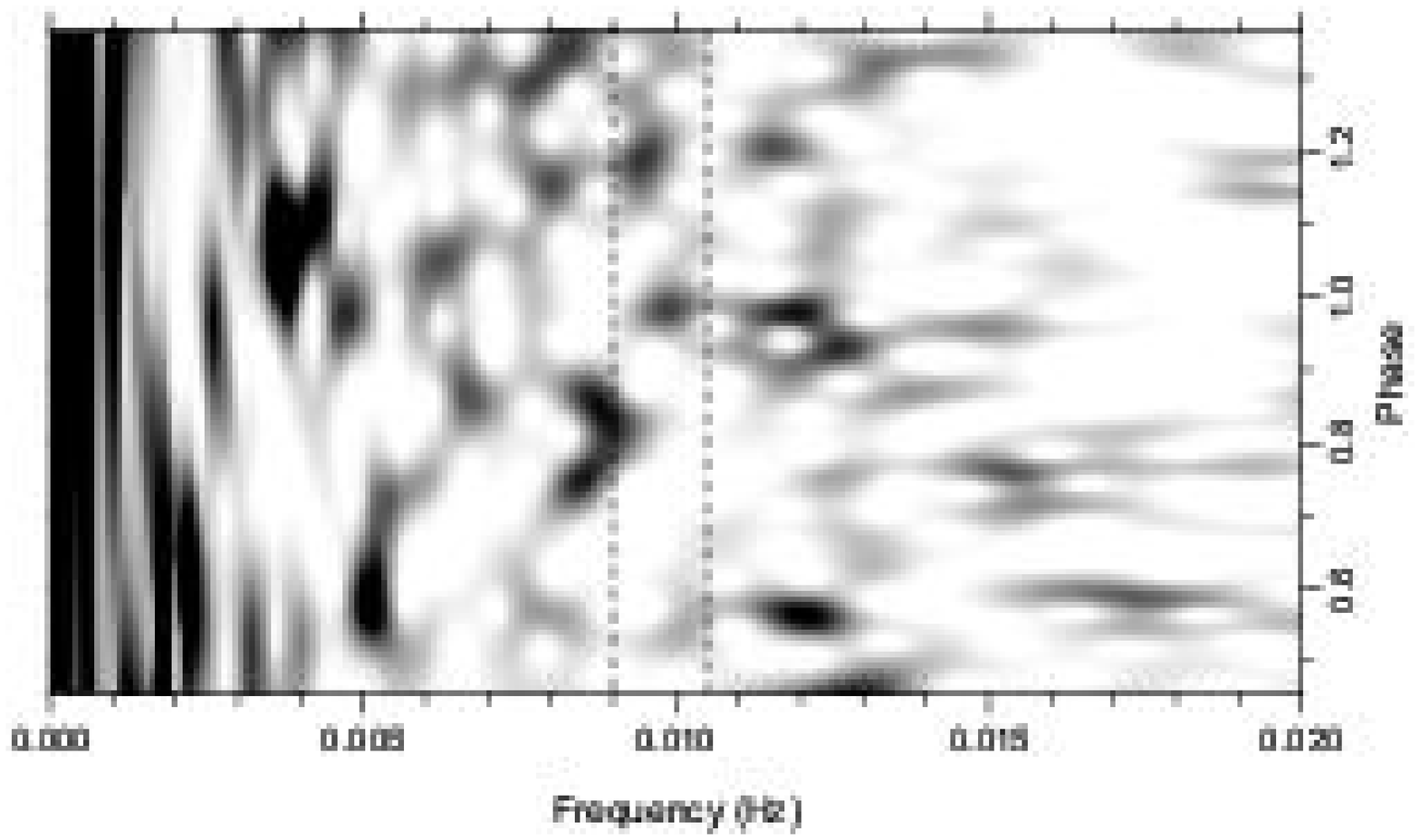}
\\
\vspace{5mm}
\includegraphics[width=99mm]{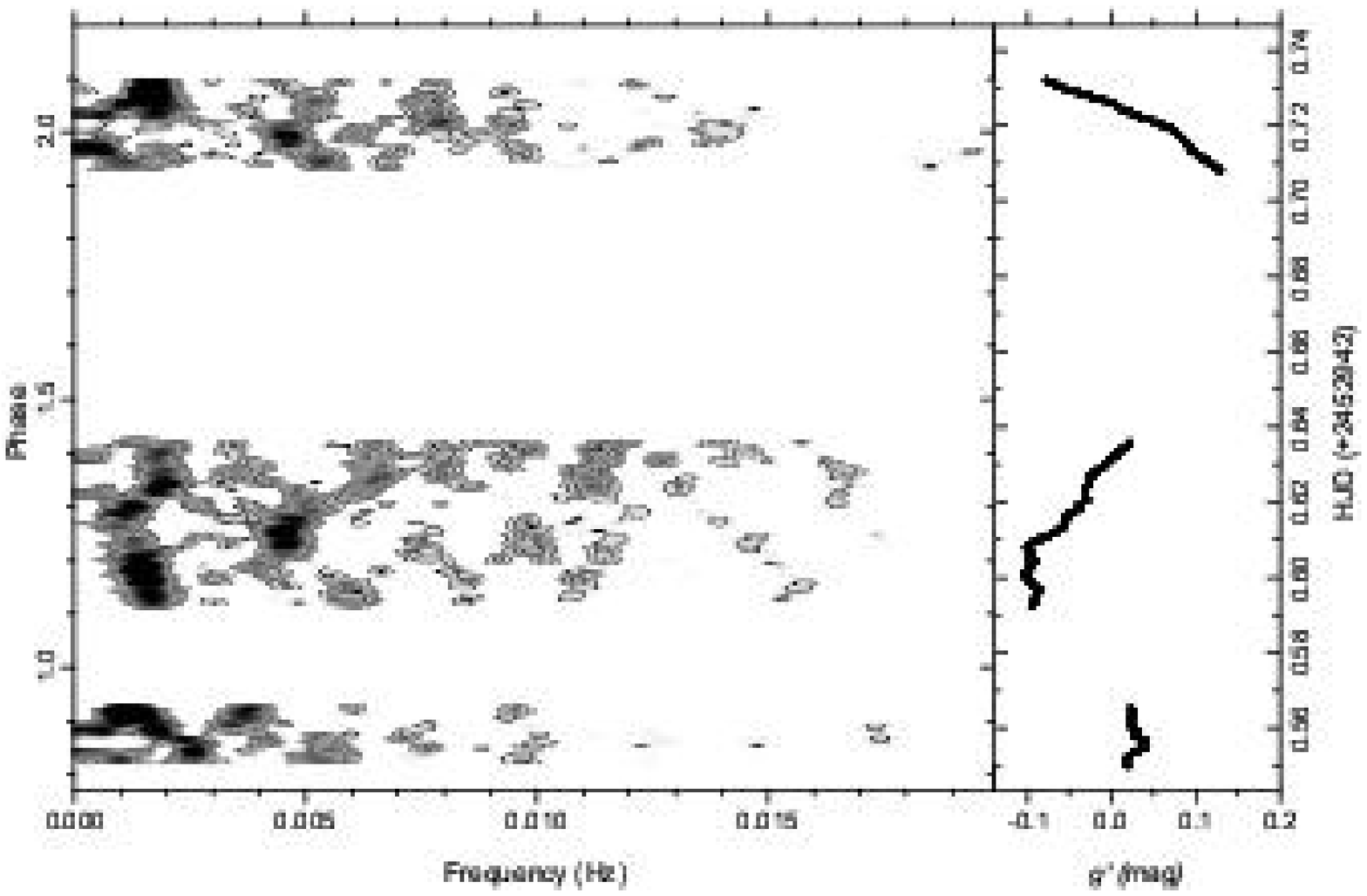}
\includegraphics[width=75mm]{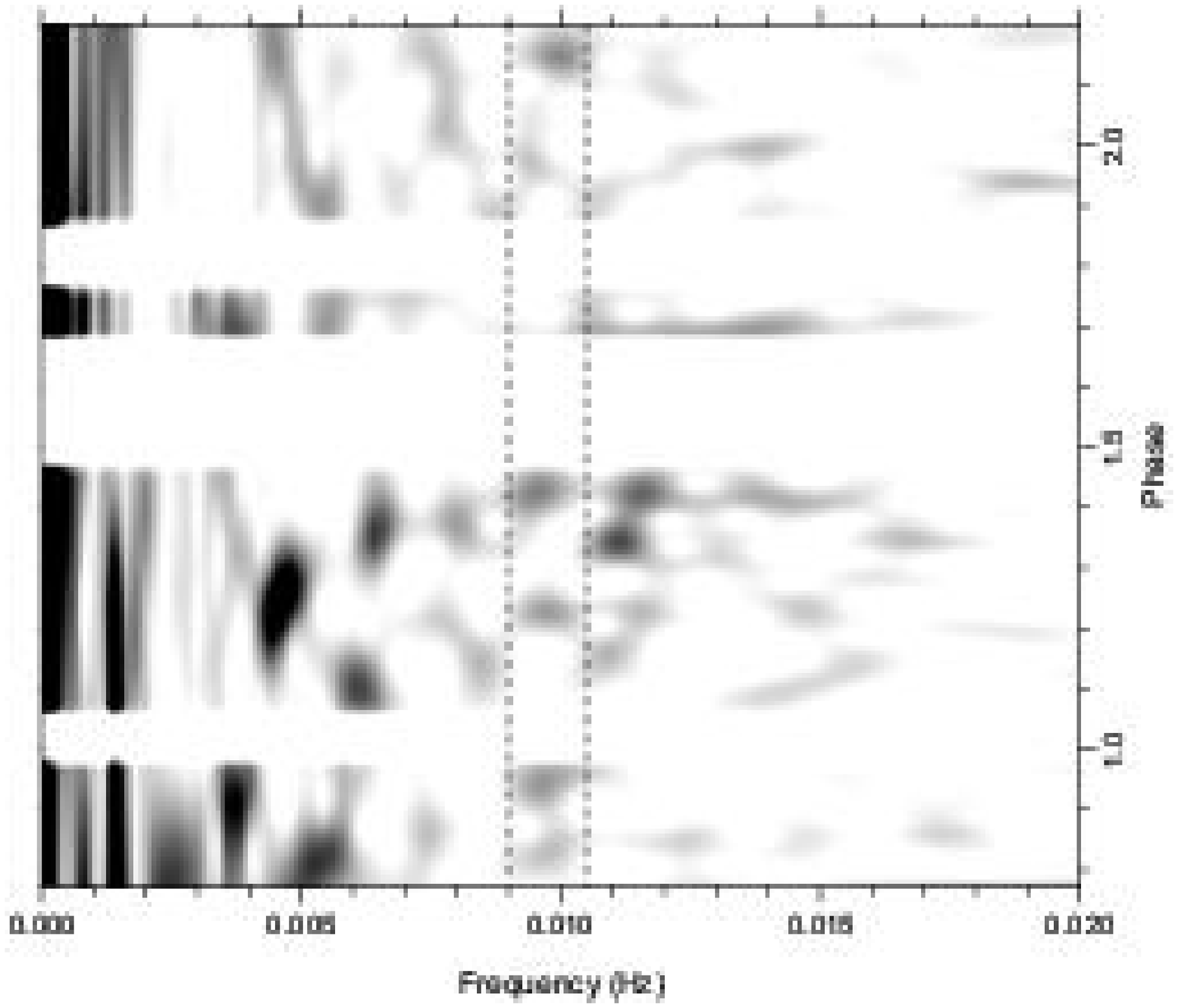}
\\
\vspace{5mm}
\includegraphics[width=99mm]{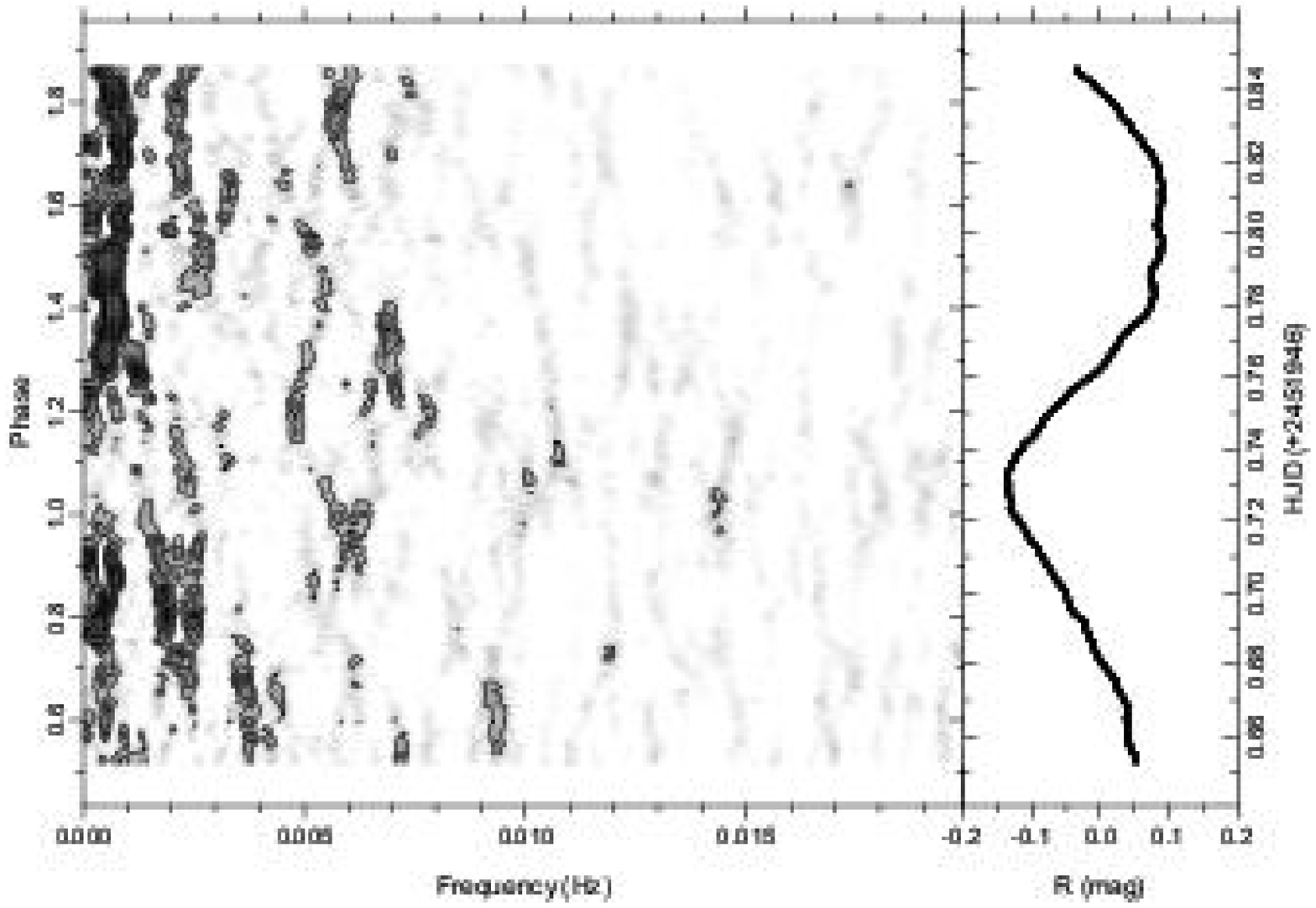}
\includegraphics[width=75mm]{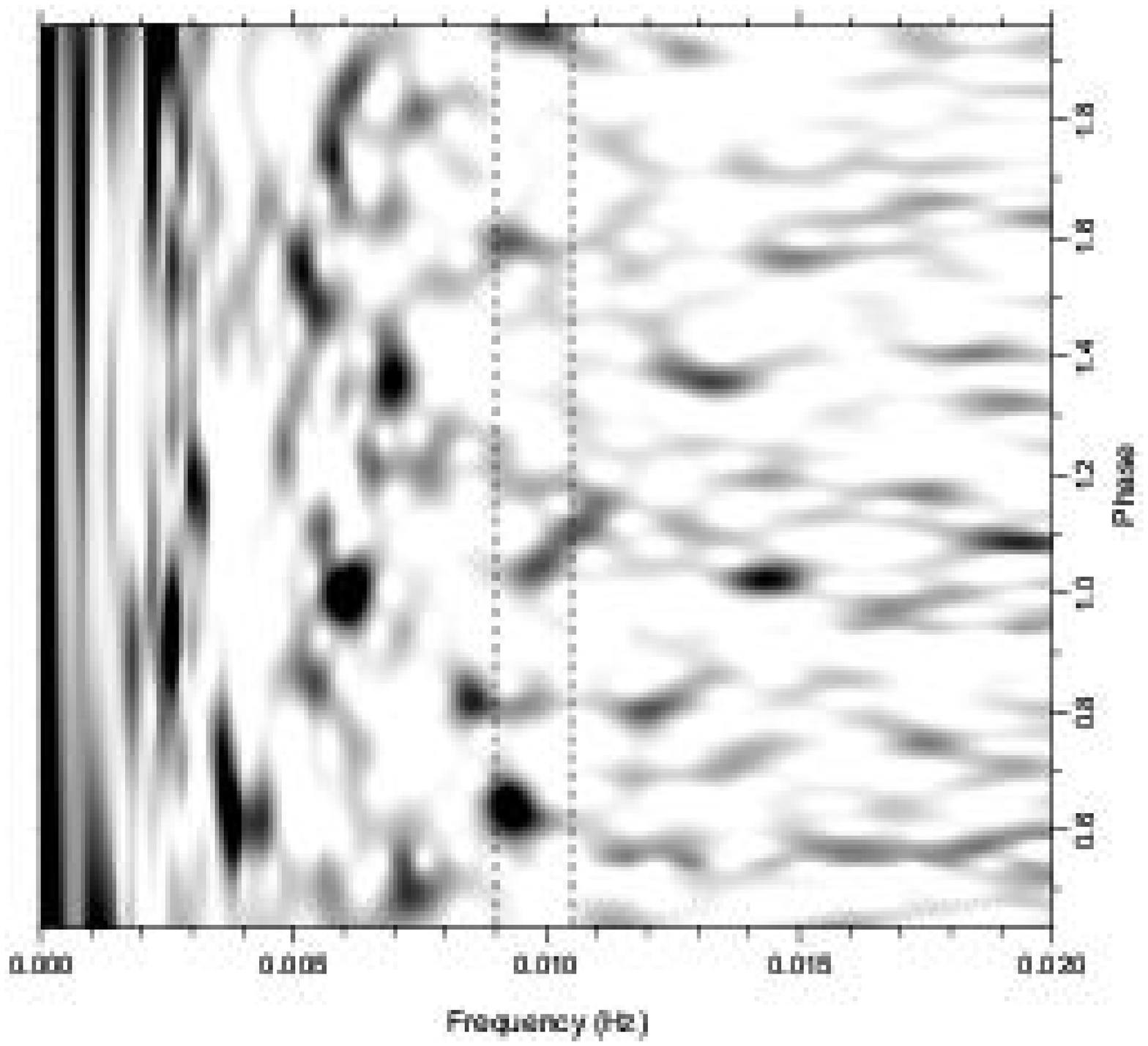}
\caption{
The sliding periodograms (slidograms) and the WWZ wavelet transforms (scalograms)
for the second night (upper row), first night (middle row) and
set 1946 (bottom row), shown as grey-scale images.
The slidograms are displayed in the left panel of the graphics
in the left column, while the right panels of these graphics show the
light curve. The scalograms are displayed in the right column.
Isophots in the sliding periodograms display the significance levels:
90\%, 99\% and 99.9\% for the second night and 50\%, 90\%, 99\% and 99.9\%
for the first night and the set 1946.
Dashed vertical lines in the scalograms images
denote the frequency range with regular appearance of ``the drops''
(see text for explanation).
Phases of \textit{photometric} period are calculated according to
the ephemeris for times of maxima:
HJD$_{max}$=$2450492.3006+0.14270785E+5.5\times10^{-10}E^2$
(Neustroev et al., in preparation).
}
\label{wavelets}
\end{figure*}

The sliding periodograms (or \textit{slidograms}) as grey-scale images
are displayed in the left
column of Fig.~\ref{wavelets}. As the slidograms are based on
the usual Lamb-Scargle periodograms we can easily calculate
the significance levels, marked in the images by isophots.
Immediately one can see that some features show  substantial frequency variations with time.

%[width=84mm]

\subsubsection{Wavelet analysis}

The WWZ method \citep{Foster} does not require time series to be evenly
spaced and without gaps, so we can work with the original data.
On the other hand, in order to find a compromise between the frequency
and temporal resolution an optimal value of the decay constant $c$
has to be chosen (in fact, $c$ is playing the role of a parameter).
After some experiments we settled on $c$=0.001, although this value is
not optimal for all frequencies.

Our results  are shown in the right column of
Fig.~\ref{wavelets}. There is a strong simularity between the slidograms and
the wavelet transforms (scalograms)  but
the latter have improving frequency resolution in the frequency range
of about 0.003 -- 0.013 Hz. At the same time, the lowest frequency
part of the scalograms is very smoothed along time axis (i.e. has
insufficient temporal resolution). The highest frequency part, on the
contrary, is smoothed along the frequency axis.

\section{Interpretation}

The main distinctive feature of the power spectra of FS Aur is the presence of
a number of strong peaks. These however, are not seen on both
nights of observations (Fig.~\ref{power}).
On the other hand, the multipeaked comb-like structure of the
periodograms testifies, most likely, to the existence and interrelation of these peaks
but not  to their stability.
Analysis of the spectra of separate subsets indicates that these periodic
processes are of short duration.

On close inspection
one can see that  most of the peaks in the power spectra are grouped
into a few groups in the following frequency ranges: 0.8--1.1~mHz,
1.3--1.8~mHz, 2--4~mHz, 4--6~mHz, and maybe a few others
with higher frequencies. Such grouping are justified from a consideration
of the slidograms and the scalograms.

We would like to note the most prominent features in these:

\begin{enumerate}

\item There are a number of lines directed along the time axis, and
practically no lines directed perpendicularly of this axis
(such lines in the scalograms can be explained by the insufficient
frequency resolution at the highest frequencies).

\item There are no strictly vertical lines existing continuously
in all time ranges.

\item Nevertheless there are a few very strong short segments of nearly
vertical lines. In the scalograms they sometimes look like ``drops''.
Some of them are located along the time axis.
It explains why the power spectra of some subsets show the
very strong peaks that are missing in the other subsets.

\item Frequently one can see splitting, or on the contrary~-- confluence,
of lines. In some areas of the scalograms this looks like a honeycomb
structure.

\item There are long and short sine-like lines with variable intensities
and frequencies. This feature explains the grouping of the peaks that
was marked earlier.
\end{enumerate}

Some of these features (especially the last) were unexpected,
so we conducted a similar analysis of the light curves of several comparison stars and found no similar peculiarities.
To confirm these features we also analyzed the light curve
of set-1946 and obtained basically the same results (see bottom panel
of Fig.~\ref{wavelets}).

\begin{figure}
\includegraphics[width=84mm]{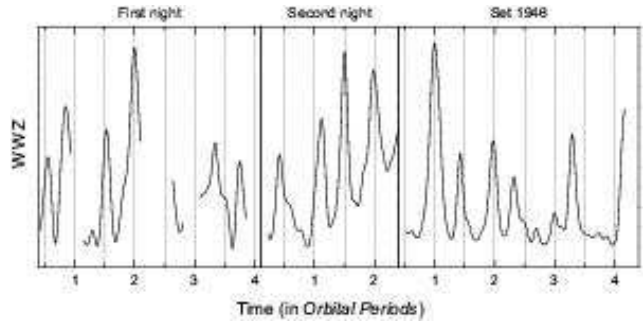}
\caption{
Time dependence of the WWZ integrated in the frequency range of 9--10.5
mHz. Time of observations has been transformed to the orbital periods with
arbitrary phase shift for display purpose.}
\label{wwz_an}
\end{figure}

Producing slidograms and scalograms from known test time-series
(see Appendix~\ref{Appendix})
the above numbered features at our time-series can be interpreted
as follows:

\begin{enumerate}

\item Orientation of the lines in the slidograms and
the scalograms mainly along the time axis indicates the
existence, in the light curve of FS~Aur,
of sine-like components. Moreover, values of all the periods obtained
in the Lomb-Scargle periodograms, are verified in the scalograms,
showing that
these periodograms have registered quasiperiodic
signals which arose and disappeared in a random way.

\item Absence of the strictly vertical lines in all the scalograms
testifies to absence of strictly periodic (coherent) signal in the
investigated frequency range.

\item However, strong pieces of nearly vertical lines seen at some
frequencies ($\sim$1.5~mHz, $\sim$3.9~mHz, probably at $\sim$13~mHz),
can be seen in all the scalograms. Due to observed significant variability
of their frequency and amplitude with time, we are not ready to call these
features strictly periodic. In fact, their behaviour resembles
Autoregressive processes (Fig.~\ref{test3wwz}) and most likely these are
some kind of classical QPOs.
But, we would like to pay particular attention to the so-called ``drops''
and short lines seen also in all sets of observations at $\sim$9.5--10 mHz.
These drops are confined to a narrower range of frequencies, and seem to
be located along the time axis at intervals of $\sim$0.2 of the photometric
period, that corresponds to nearly half of the orbital period.
We have integrated the WWZ in the frequency range of 9--10.5 mHz,
and transformed the time of
observations to the orbital periods, with arbitrary phase shift for every
dataset (Fig.~\ref{wwz_an}). One can see the WWZ maxima are really located
through every half of the orbital period, with a few exceptions in the first
night and set 1946. These exceptions, as well as scattering of the drops around
the mean frequency, can be easily explained by an influence of noise
(see an appearance of the test function $F_5$ in the slidogram in our last
test in Appendix~\ref{Appendix}).
This means that we probably see the recurrent appearance of modulations,
seen every half of the orbital period.

\item The honeycomb structure of the scalogram is a corollary of strictly
stochastic processes similar to white noise (Fig.~\ref{test2wwz}).
The splitting and confluence of the strong lines can arise also due to
the Autoregressive processes.

\item
Sine-like lines can arise from strictly periodic processes
similar to an frequency modulation, and from Autoregressive processes.
However, in the latter case it is difficult to obtain a sensible
frequency deviation and regular sinusoids.
In this context a sinusoid at $\sim$5--6~mHz
is apparent. This can be seen in all the sliding periodograms
and scalograms, though not during all time of the observations.
It is interesting to note, that in the second night and possibly in set-1946
this sinusoid is accompanied by another with higher frequency, where their
maximal strength is reached in an antiphase.

\end{enumerate}

\section{Discussion}

With the exception of the
205.5-minute periodicity, the main characteristic of variability of
FS Aur is the usual CV flickering plus QPOs. The two following
features deserve special attention: (i) the frequency-modulated oscillations
at $\sim$5-6 mHz (150-200 sec);
and (ii) the recurrent appearance of modulations with period
of $\sim$100 sec. The former  is simply extremely unusual and to the
best of our knowledge has never been observed before, whereas the second
could be the spin period of the white dwarf or at least connected to it.

\subsection{Significance tests}
\label{red-noise}

At low frequencies the usual statistical significance tests,
based on the assumption of Gaussian white noise, should not be applied to
determination of significance levels of peaks in Lomb-Scargle
periodograms. Red-noise more adequately to describes CV flickering.
In contrast to white noise, the power spectra of red-noise
shows a continuous decrease of spectral amplitude with increasing frequency,
and frequently shows very strong peaks.

We have investigated the significance of the observed power-series using two methods.
Firstly, we have calculated
confidence limits for our observed power spectra and shown which peaks
are really "significant". However it is well known that some strictly
periodic functions can produce quite weak peaks in power spectrum, even
without any noise (see examples in Appendix~\ref{Appendix}). So,
in the second test, we calculated a set of artificial scalograms and
compared them with the observed ones in order to understand if red noise could
reproduce the observed features.

We calculated the confidence limits using a well developed approach from the
Geosciences (\citealt{Schulz}). Briefly this procedure can be described as follows.
Firstly the observed time-series were approximated by Autoregressive (AR) processes
of different order, in order to determine which
AR model reproduces the power spectra slope. These derived optimal parameters are then used
for the calculation of a large set of simulated time-series using  the observed times.
These artificial time-series can then be used
to estimate the significance levels of the Lomb-Scargle periodograms
(for more details see also paper of \citet{Hakala} for  application of this
method to astronomical data analysis).

Estimation of the AR parameters from the time-series is
a relatively straightforward procedure, but only for evenly sampled
data. For unevenly spaced data this technique would require
some sort of interpolation. Unfortunately, this procedure may result
in a significant bias \citep{Schulz2}. \citet{Schulz} presented
an algorithm  that estimates the AR parameters directly from unevenly
spaced series, but only for AR process of the first order.
This is probably insufficient to describe our time-series\footnote{\citet{Hakala} used
AR(7) model to reproduce the power spectrum slope of YZ Cnc correctly.}.
Fortunately, our second night's time-series are almost evenly spaced
and contain only short gaps, and consequently should not affect the power
spectra continuum distribution.

\begin{figure}
\includegraphics[width=84mm]{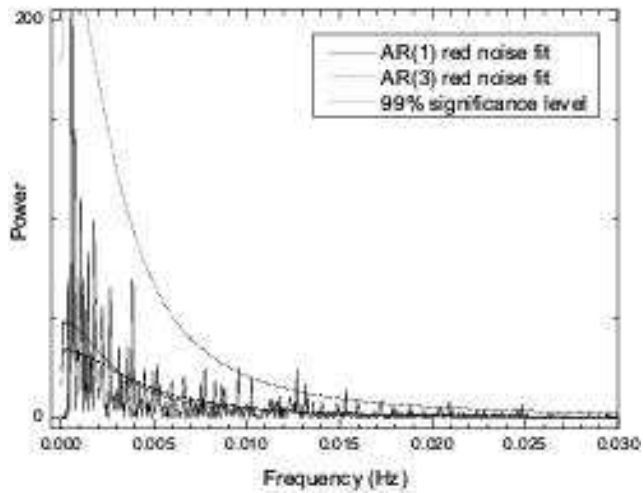}
\caption{
The $g'$-band power spectrum of the second night with AR(1) and AR(3) red
noise models, and the 99\% significance level based on AR(1).}
\label{rednoise}
\end{figure}

Because the light curve of FS Aur  includes a periodic component
with the period of $\sim205.5$ min, the time-series has been pre-whitened
before the subsequent analysis.
Fitting AR processes to the time-series of the second night shows
that already a third-order AR model is more than enough to reproduce
the observed power spectrum slope. However, there is insignificant
difference between AR(1) and AR(3) in the high-frequency part of the spectrum,
AR(1) is better in the medium-frequencies. So we have decided that
a first-order Autoregressive process is sufficient to describe
the red-noise flickering in FS Aur. Using this AR(1) model we have calculated
the significance levels for the power spectrum of the second night and shown
these in Fig.~\ref{rednoise}.
One can see that almost all peaks are below the 99\% significance
level and can be easily explained by red noise confirming our previous
conclusions. At the same time it says nothing about the existence of the
frequency-modulated oscillations and 100 sec variability as we did not expect
to have strong peaks from them. However, the latter shows a  significance level of more than 99\%.

In order to investigate more throughly the significance of these results
we performed a Monte-Carlo analysis.
On the base of earlier obtained AR(1) model we have simulated more than 200
time-series, which were used for calculating the scalograms.
We then performed a quick-analysis of every artificial scalogram
in the same way as the observed ones. Our purpose was to find
structures similar to sinusoids on $\sim$ 5-6 mHz and the recurrent
``drops'' in all the frequency range up to 20 mHz, not only near
to 10 mHz. Worrying about the clarity of experiment we have
also repeated this analysis for more than 200 artificial scalograms obtained
on the base of AR(3) model.

\begin{figure*}
\includegraphics[width=160mm]{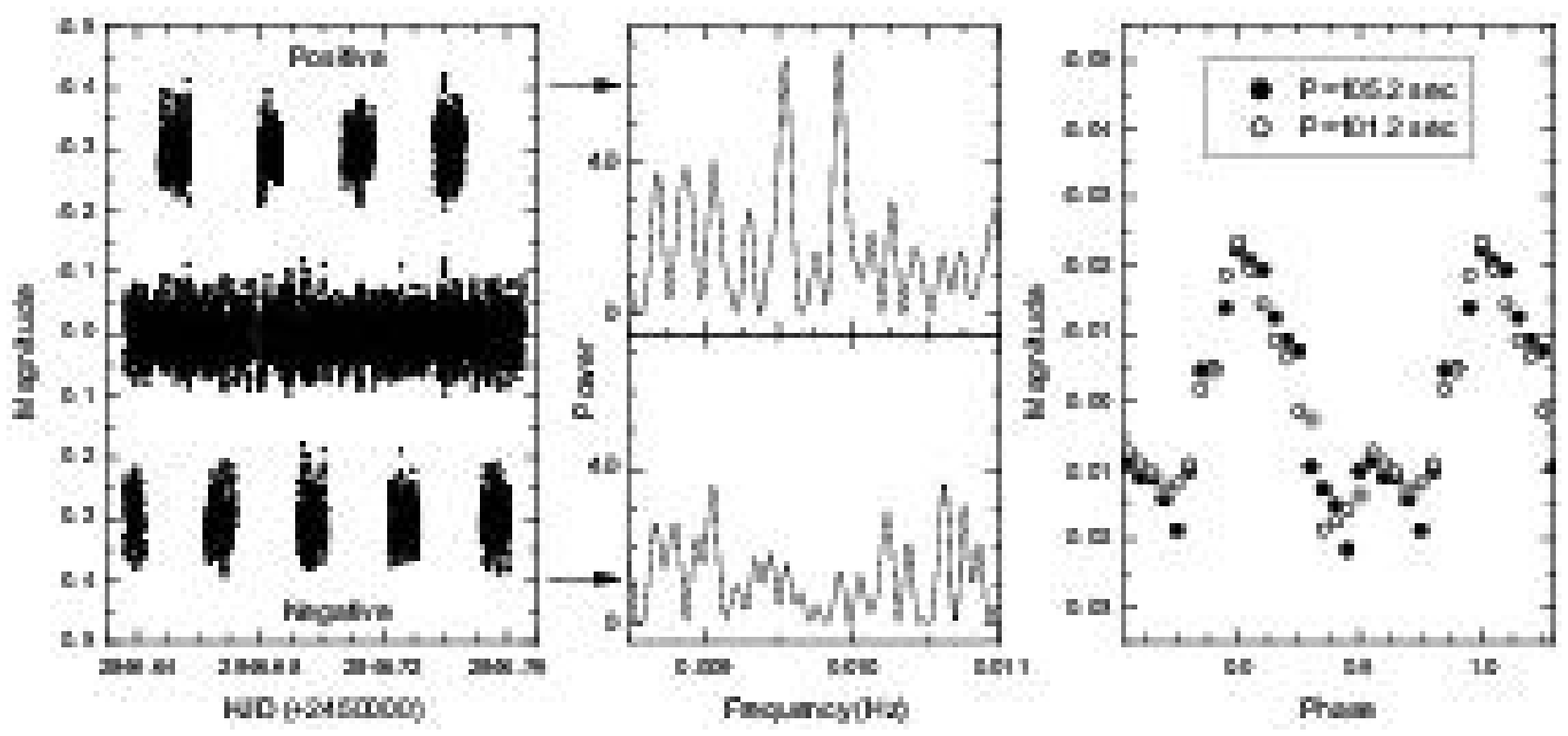}
\includegraphics[width=160mm]{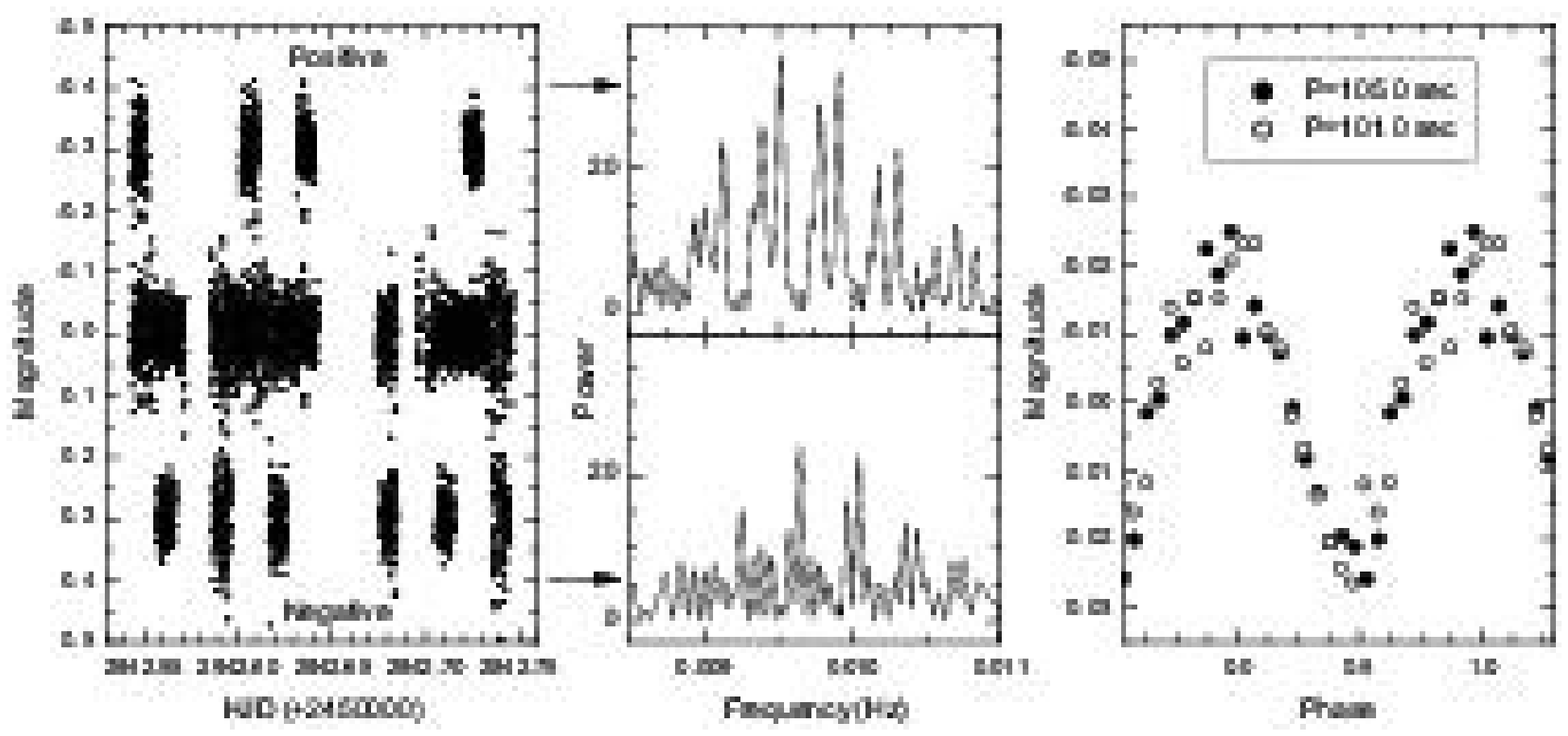}
\caption{
The upper row belongs to the second night's data, the bottom row to the first
night's data.
Left frame: Detrended $g'$-band light curve (middle) and
the Positive- (upper) and the Negative- (bottom) samples of it (see text for explanation).
The samples are displaced by -0.3 and +0.3 mag.
Middle frame: Power spectra of the samples.
Note that frequencies of the two strongest peaks in the power spectra of
the Positive-samples of the both nights completely coincide.
Right frame: The mean waveform obtained by folding the Positive-sample on the period
of $\sim$101 and $\sim$105 sec.}
\label{power100}
\end{figure*}

We found that (i) none of the scalograms showed  clear
large-amplitude sinusoids; (ii) in almost 500 trials we performed to search
for series of periodicities separated by half the orbital period,
this was only observed twice in the simulated data in which only
three successive periodicities (with periods of $\sim70$ sec and $\sim100$ sec)
were observed. Whereas in \textit{all}
the real data sets this sequence was observed at least \textit{four} times
(Figs.~\ref{wavelets} and \ref{wwz_an}).

So our conclusion is that both features (the frequency-modulated oscillations
and the recurrent appearance of modulations with period of $\sim$100 sec) are
real and cannot be readily explained by red-noise.

\subsection{100 sec variability}

The most persistent peaks of high significance levels are those
concentrated around frequencies corresponding to $\sim$0.01 Hz. However,
we were unable to find any coherent modulations with periods of
50--100 sec existing continuously in all time ranges, as well as with other
periods shorter than tens of minutes. This is disappointing, but probably
not very surprising. Even within the suggested hypothesis of the rapidly
rotating and precessing magnetic white dwarf one would expect to detect
the emission from the accretion poles in high energy range only, and to
see only the reprocessed/reflected echo in optical wavelengths. If the
high energy beam hits different parts of the disc due to the precession,
then magnitude of pulses can be variable or they might disappear
altogether during part of the period. Therefore we have reasons to think
that detected oscillations with the period of about 100 sec, seen during
short time in the slidograms and scalograms every half of the orbital period,
can turn out to be the required variability.

\textit{Are these real effects or noise?}
One can see several other series of ``drops'' located along the time
axis of all the scalograms (for example, at $\sim$12--13 mHz). However, in all these
cases the time interval between ``drops'' is variable, and only for 100-second
oscillations is this time interval approximately constant and equal to half of
the orbital period. Last fact is especially indicative. Any discovered periodic
events with the period equal to the orbital one or multiple to it, should be
considered very seriously.

In previous subsection we have already shown that there is only a small probability
that these 100-seconds events are due to red noise.
Additional evidence for their reality come from an assumption that the 100-second
oscillations can be detected \textit{periodically} and only for a \textit{short} time.
Using the detrended time-series of the second night we have prepared two samples,
each of them includes about a third of all points.
These samples consist of subsets located through each half of the orbital period.
The subsets of the first sample are centred on the ``drops'' in
the scalograms (which we will refer to as ``the Positive-sample'')
while the subsets of the second sample (``the Negative-sample'') are shifted relative
to them on a quarter of the orbital period (Fig.~\ref{power100}, upper-left frame).
If our assumption
is correct then one can expect to see stronger $\sim$100-seconds peak in the
power spectrum of the Positive-sample in comparison to the Negative-sample.
Our analysis confirms this (Fig.~\ref{power100}, upper-middle frame).
The Positive power spectrum shows two strong peaks (at $\sim$9.5 and $\sim$9.9 mHz)
completely absent in the Negative spectrum. We have also applied the same approach
to the first night's dataset. Though the spectral window in this case is much
more complex, we have obtained exactly the same results (Fig.~\ref{power100}, bottom row).
Moreover, frequencies of the two strongest peaks in the power spectra of the
Positive-samples of the both nights completely coincide.

On the basis of these data sets we are unable to determine which of these peaks
is the real period. Either one of them could be the alias of other.
Taking into account the equal strength of the peaks, also both periods could be present
in the time-series.
Right frames of Fig.~\ref{power100} show the mean waveforms, obtained by folding the
data on these periods (101.0 and 105.0 sec for the first night, and 101.2 and 105.2 sec
for the second). All the waveforms are very similar and almost symmetric.
The full amplitude is $\sim$0.04 mag.

Our analysis shows that the $\sim$100-seconds oscillations
really exist in the light curves of FS Aur. Thus our hypothesis concerning a
precessed rapidly rotating white dwarf in FS Aur still remains open and
has received observation confirmation, if these oscillations are
connected to the spin period.
Nevertheless, new high Signal-to-Noise ratio observations with high time
resolution are now required. We suggest that X-ray observations would be
very useful for confirming of our results.

\subsection{Frequency modulation}

The detected frequency-modulated oscillations are real as we observed
them at different times with different instruments. Moreover, as it
follows from results of the analysis in Subsection~\ref{red-noise},
they are probably not due to red noise. Unfortunately, we can say nothing
concerning such extremely interesting variability in other stars
inasmuch as, to the best of our knowledge, a similar analysis of a CVs
photometry has been rarely performed \citep{Warner-Brickhill},
and classical methods are unsuitable to detect it.
For the time being we have no plausible explanation for these modulations.

\section{Conclusion}

In this paper we presented the analysis of high-speed photometric observations
of FS\,Aur taken with high-speed camera ULTRACAM in late 2003.
These observations were intended to search for variability in 50--100 second
time domain consistent with the rotation period of a white dwarf. Observing such
a periodicity would help explain the
coherent photometric period of 205.5 minutes which exceeds the
spectroscopic period of 85 minutes.

Using various methods, including wavelet-analysis, we found that
in the  frequency range above 0.001 Hz FS~Aur behaves in a similar manner
to the other dwarf novae and the principal source of variability is
flickering, commonly observed in these systems.
In addition,  we detected oscillations with the period(s) of $\sim$101 and/or $\sim$105 sec,
observable for a short time for half of the orbital period.
These oscillations may be associated  with the spin period
of the white dwarf, strengthening our hypothesis on the precessing
rapidly rotating white dwarf in FS Aur.

\section*{Acknowledgments}

VN acknowledges support of IRCSET under their basic research programme and
the support of the HEA funded CosmoGrid project.
Wavelet analysis was performed using the computer program WWZ,
developed by the American Association of Variable Star Observers.
We thank Tom Marsh and Vik Dhillon for their observational support and
valuable suggestions improving the manuscript.
The paper has benefited from constructive comments
by the referee.

\appendix

\section{Wavelet analysis of model functions}
\label{Appendix}

In order to understand the results presented in this paper we have
analysed synthetic function of a similar form to those observed. Our
approach has been to calculate synthetic time-series and investigate
their scalograms in the presence of noise. In particular we were
interested in how their wavelet transform performs in
the presence of noise. The wavelet transform in this context has also
been described by  \citet{Szatmary} and \citet{Vityazev}. Our intention
was to determine the slidograms and scalograms for synthetic functions
and compare these with the strong features of the observed datasets. How
noise effected the behaviour of these was of interest as was the effect
of decay constants, e.g. the $c$ term in Foster's method.

\begin{figure}
\includegraphics[width=84mm]{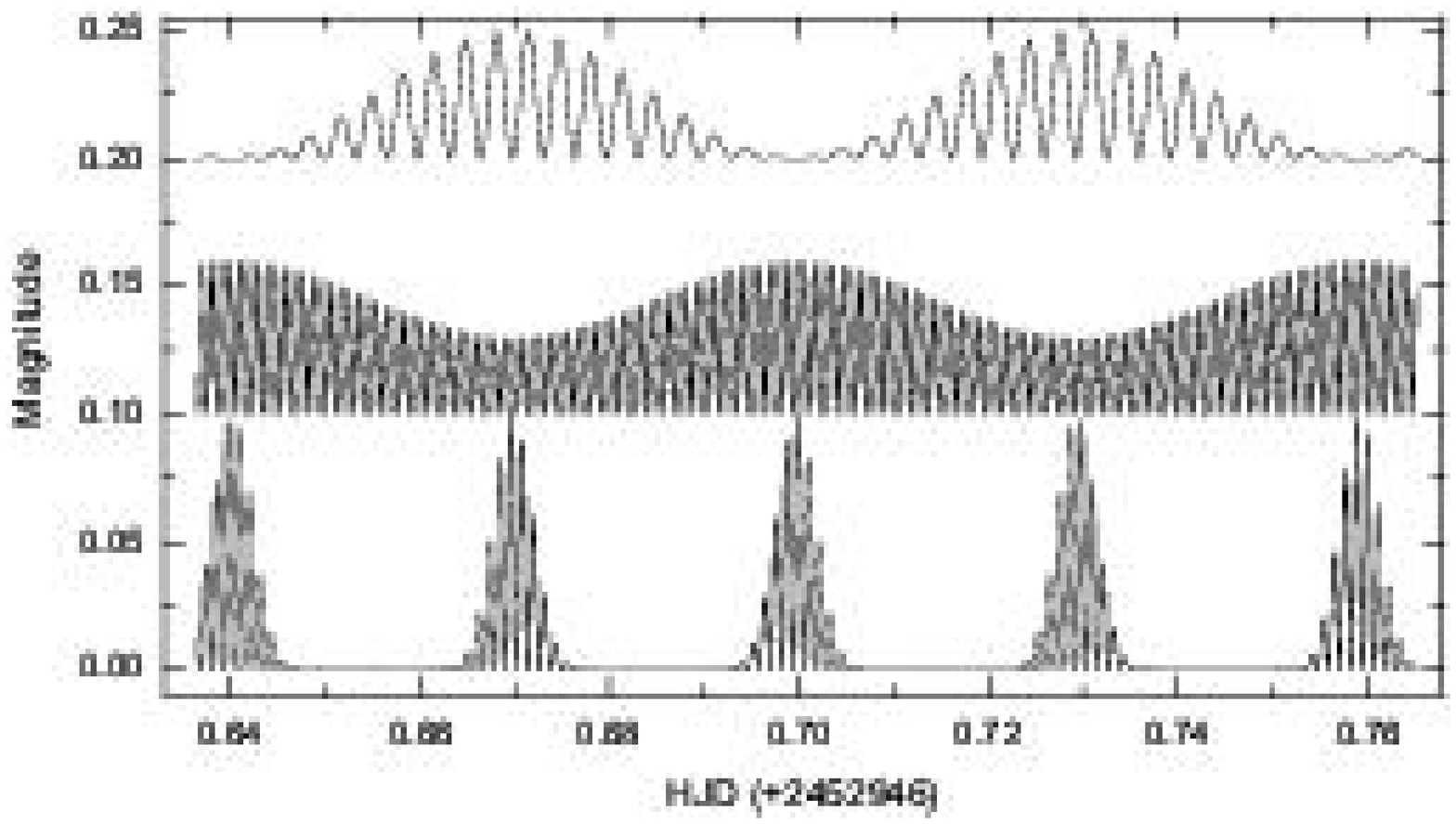}
\caption{
Plots of the test functions $F_3$ (upper), $F_4$ (middle) and $F_5$
(bottom), shifted by 0.1 in the y-direction.}
\label{testfunc}
\end{figure}

Our model scalograms were generating using the same sampling that
we had on our second night of observation.
For the first test we generated
artificial time-series by adding seven simulated signals using the
following functions:

\begin{enumerate}

\item The simple sinusoid $F_1$ with constant semi-amplitude of
0.05 and frequency of $\nu_1=0.002$\,Hz.

\item The sinusoid $F_2$ with constant semi-amplitude of 0.05,
which frequency changes three times by a saltation (0.0145\,Hz,
0.0150\,Hz and 0.0155\,Hz) at regular time intervals.

\item Two amplitude-modulated sinusoids $F_3$ and $F_4$ with frequencies
$\nu_3$ and $\nu_4$ of 0.0035\,Hz and 0.012\,Hz respectively, and with
frequency
of amplitude modulations of $\nu_{orb}=1/P_{orb}=1.94\cdot10^{-4}$ Hz.
The amplitude of $F_3$ varies sinusoidally from 0 up to 0.05, while
the amplitude of $F_4$ from 0.03 up to 0.06 (Fig.~\ref{testfunc}):

\begin{equation}
     F_3= \frac{0.05}{4}(1+\sin 2\pi\nu_{orb}t)(1+\sin 2\pi\nu_4t)
\end{equation}

\begin{equation}
     F_4= \frac{1}{2}(0.045+0.015\sin 2\pi\nu_{orb}t)(1+\sin 2\pi\nu_5t)
\end{equation}

\item One more amplitude-modulated sinusoid $F_5$ with frequency
$\nu_5$ of 0.010\,Hz and with an amplitude modulation frequency
of 2~$\nu_{orb}=2/P_{orb}=3.89\cdot10^{-4}$\,Hz. Unlike the previous
functions this sinusoid has visible oscillations (from 0 up to 0.10)
for about 33\% of the time (Fig.~\ref{testfunc}).

We calculated it, using the following formula:

\begin{equation}
     F_5= 0.05(\sin^{20}{2\pi\nu_{orb}t})\cdot(1+\sin{2\pi\nu_6t})
\end{equation}

\item Two frequency-modulated sinusoids $F_6$ and $F_7$ with carrier
frequencies $\nu_c$ of 0.0055\,Hz and 0.0080\,Hz respectively, and
with modulating frequency $\nu_m$ of $1.62\cdot10^{-4}$\,Hz=$2/P_{phot}$
and $1.94\cdot10^{-4}$\,Hz=$1/P_{orb}$, respectively.
The modulation index $\delta$ has been chosen so to obtain
a maximum frequency deviation of 0.0008\,Hz.
The amplitude of both sinusoids are 0.05:

\begin{equation}
     F_{6,7}= 0.05\sin(2\pi\nu_ct+\delta\sin{2\pi\nu_mt})
\end{equation}

\end{enumerate}

\begin{figure}
\includegraphics[width=84mm]{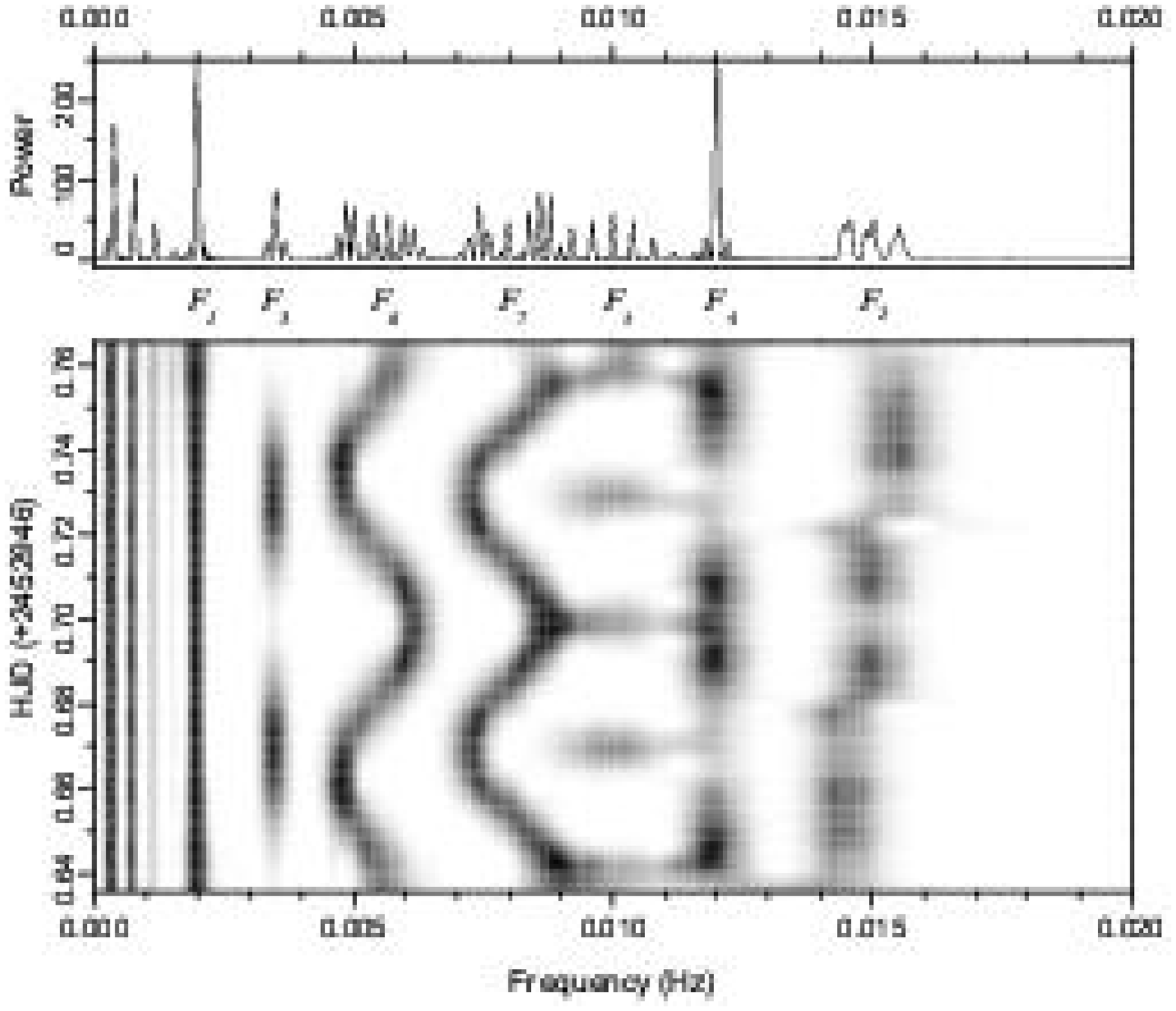}
\caption{
The Lomb-Scargle power spectrum (upper panel) and the scalogram
(bottom panel) of the first artificial time-series. Frequency
position of test functions are shown by $F_1$,...,$F_7$
(see text for explanation).}
\label{test1wwz}
%\end{figure}
\vspace{0.5cm}
%\begin{figure}
\includegraphics[width=84mm]{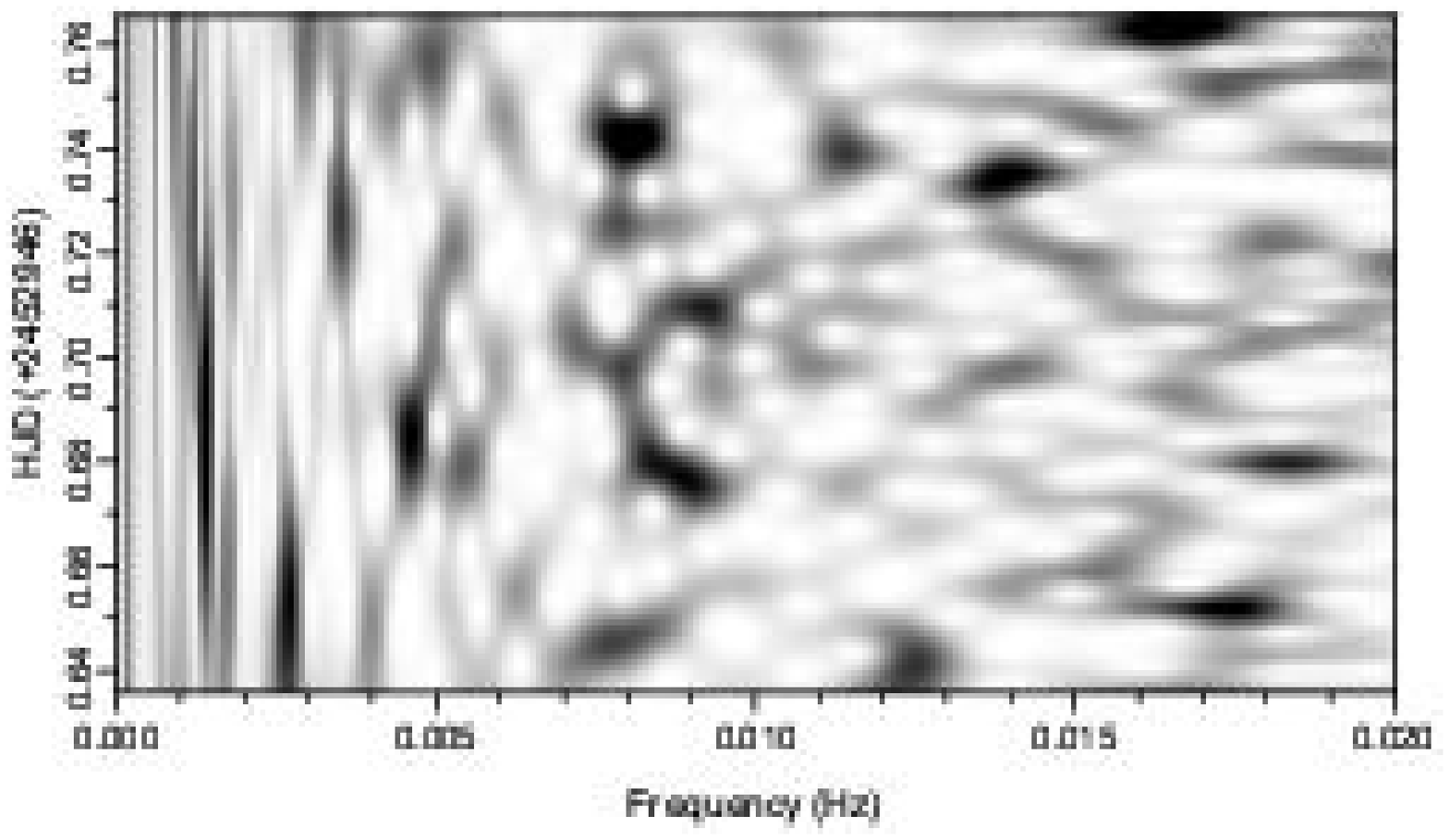}
\caption{
The scalogram of white noise.}
\label{test2wwz}
%\end{figure}
\vspace{0.5cm}
%\begin{figure}
\includegraphics[width=84mm]{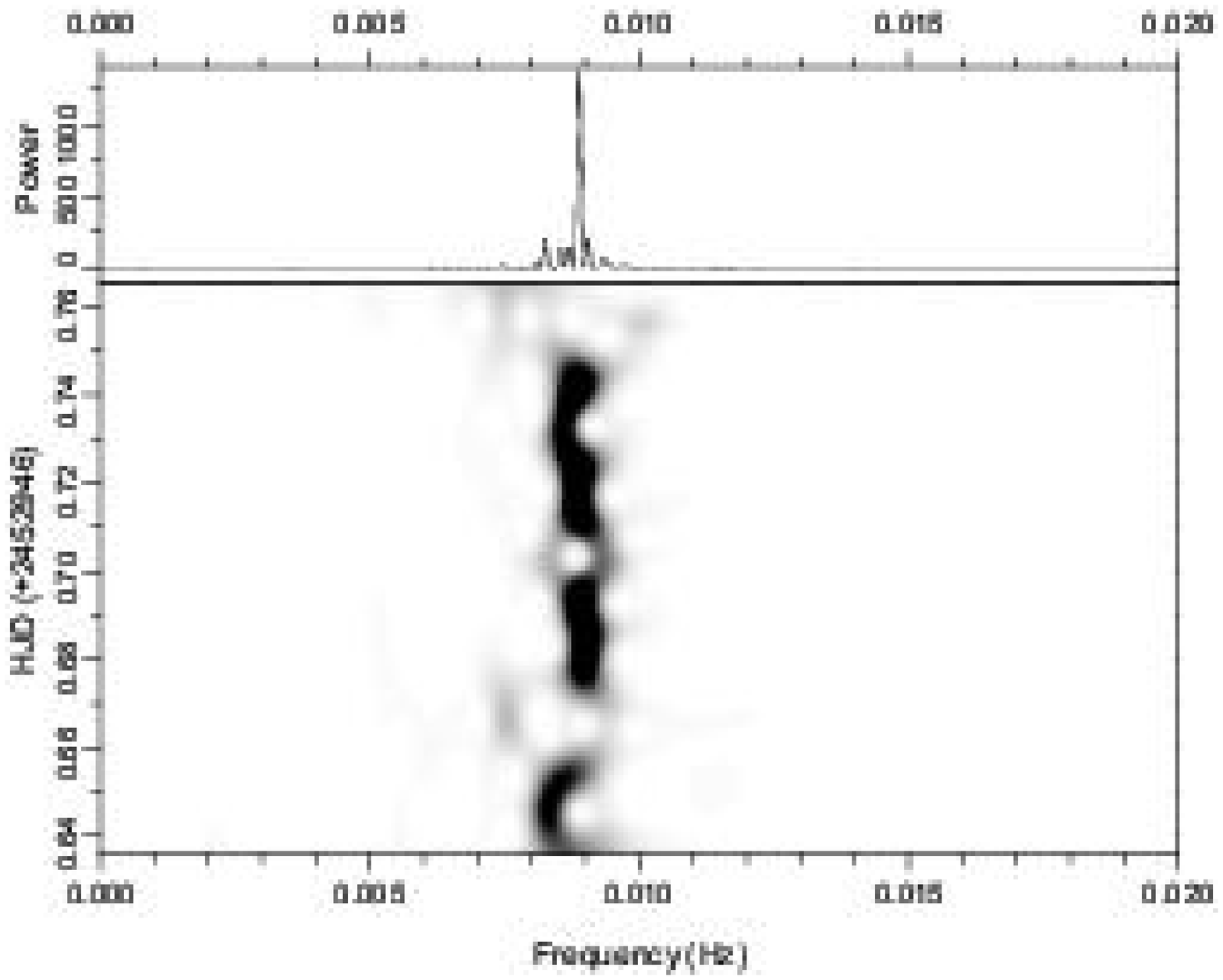}
\caption{
The Lomb-Scargle power spectrum (upper panel) and the scalogram
(bottom panel) of the AR(2) model with constants $a_1$=1.980 and
$a_2$=--0.995.}
\label{test3wwz}
\end{figure}

Figure~\ref{test1wwz} shows the scalogram and the Lomb-Scargle power
spectrum for the given artificial time-series. From this we can easily
detect the basic characteristics of all seven input signals - frequency,
duration, amplitude or frequency modulation, etc. In particular, a
difference in the modulating frequencies $\nu_m$ of functions $F_6$ and
$F_7$ is evident. However, we note a significant edge effect and
reallocation of power along frequencies between various signals. As an
example of the latter see that the brightness reduction of the vertical
lines at frequencies of 0.0145--0.0155\,Hz (function $F_2$) correlate with
spots at 0.010\,Hz (function $F_5$), as well as with different brightness
of these spots along the time axis.

The first test time-series was without added noise. This was to determine
the basic wavelet-transforms of the test functions. However the influence
of  noise on the scalograms is also of significant interest.
Figure~\ref{test2wwz} shows the scalogram of discrete white noise having a
normal distribution. Main distinctive feature of this scalogram is a
honeycomb structure,  as well as the concentrations of power, randomly
distributed over scalogram. Though the latter have relatively low
significance.

For an investigation of many processes interjacent between random
and strictly periodic, \textit{Autoregressive Models} AR($p$) can be
used
\citep{ScargleAR}:

\begin{equation}
     x_k= a_1 x_{k-1}+a_2 x_{k-2}+...+a_p x_{k-p}+\xi_k
\end{equation}

\noindent where $p$ is a model order, $a_1,a_2,...,a_p$ are model
constants and $\xi_k$ is an uncorrelated white noise. As AR(2) models can
have a quasi-sinusoidal appearance, they are particularly useful for
representing quasi-periodic oscillations in accreting binaries. Our third
test is devoted to obtaining the distinctive features of scalograms of
AR(2) processes. For this we generated time-series for an AR(2) model with
constants $a_1=1.980$ and $a_2=-0.995$. This model produces a narrow and
very strong peak in its power spectrum (Fig.~\ref{test3wwz}, upper panel)
that can be mistakenly interpreted as an indication of coherent
oscillations. However, wavelet analysis shows that it is definitely not. A
strong line seen in the scalogram (Fig.~\ref{test3wwz}, bottom panel), is
not strictly vertical, though it is directed along the time axis. It can
have frequency drifts, long enough sine-like parts, gaps, splits and
confluences.

\begin{figure}
\includegraphics[width=84mm]{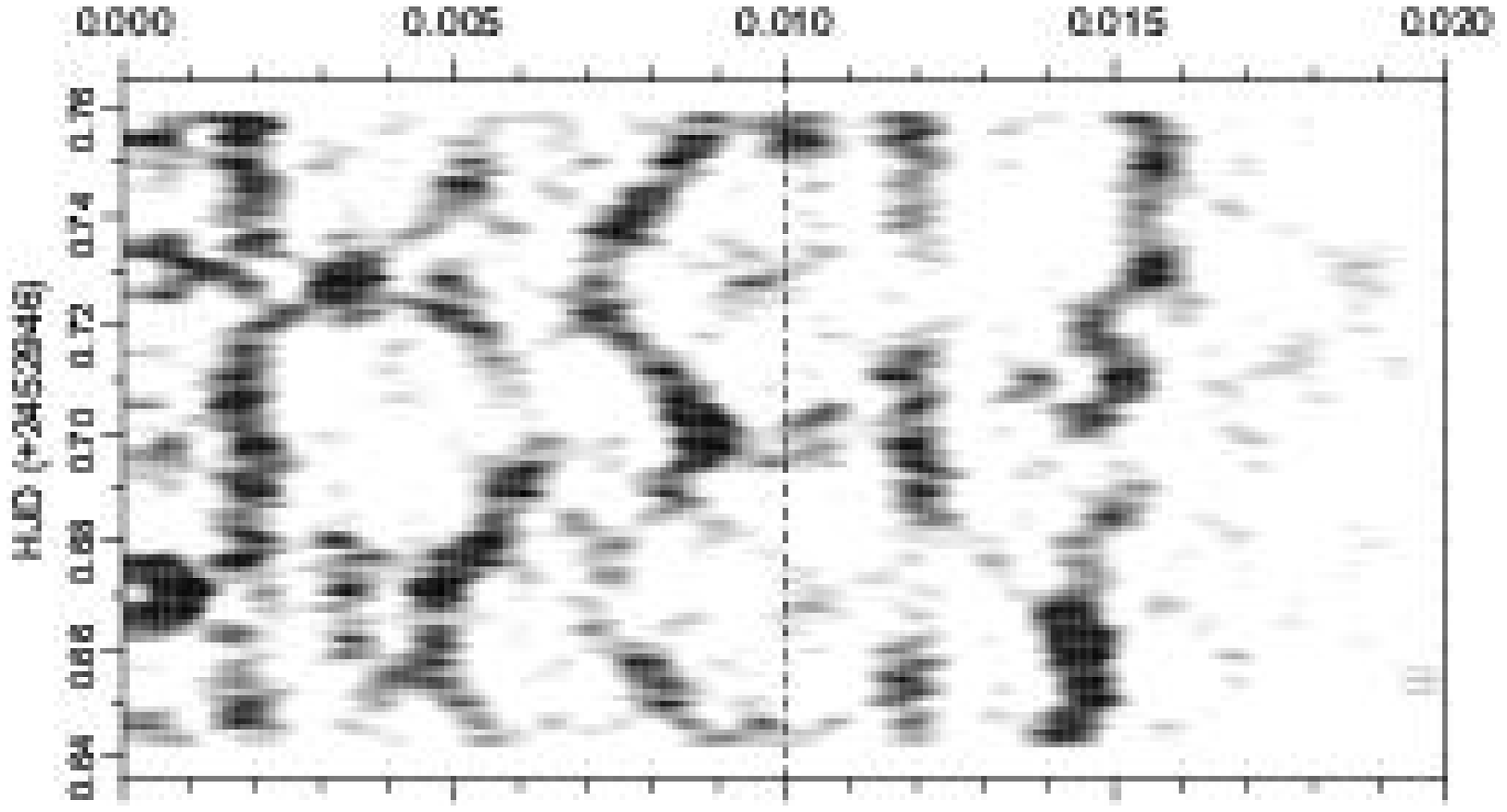} \\
\includegraphics[width=84mm]{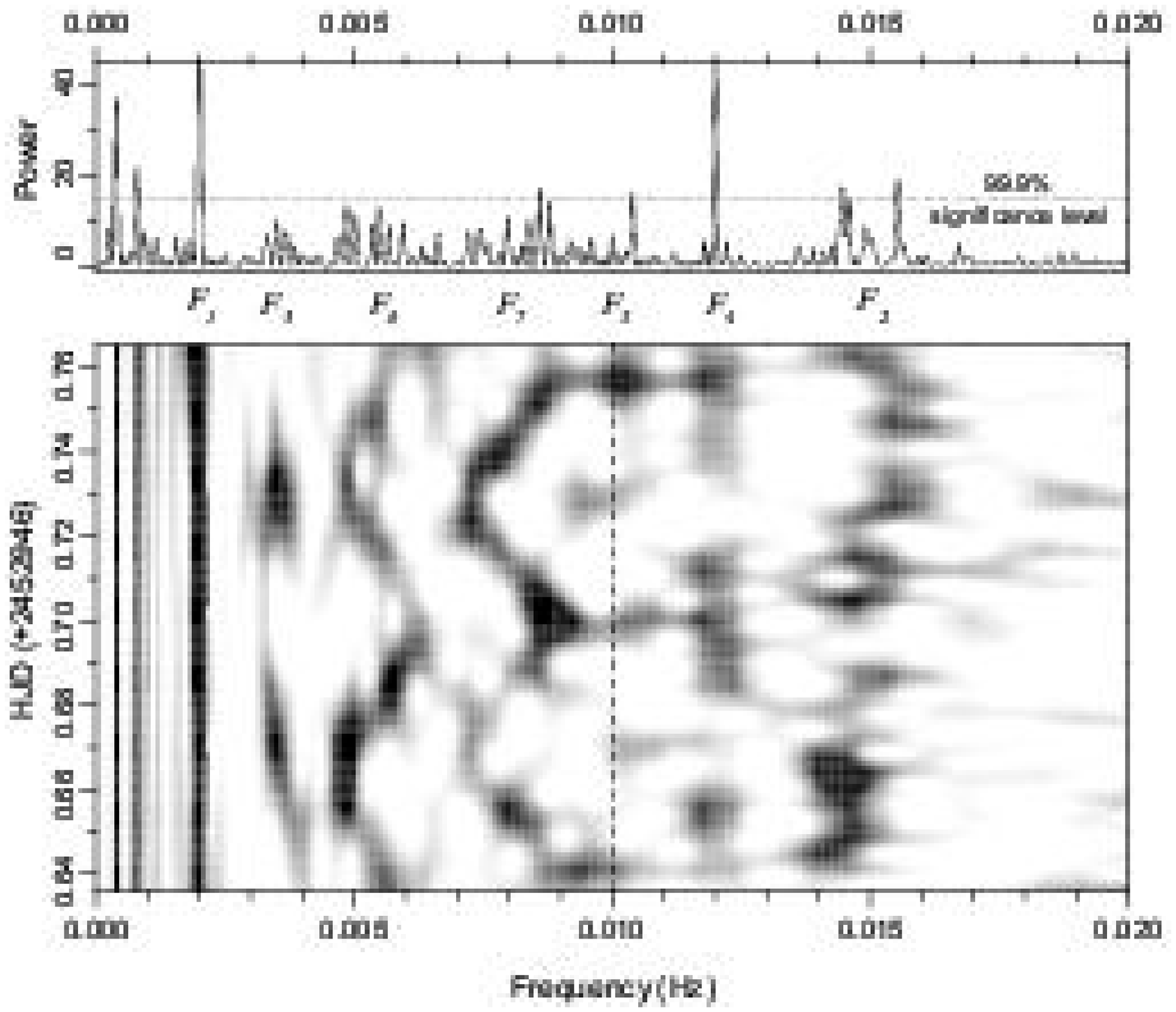}
\caption{
The slidogram (upper), the Lomb-Scargle power spectrum (middle) and
the scalogram (bottom panel) of the first artificial time-series with
added a white-Gaussian noise.}
\label{test4wwz}
\end{figure}

Finally we investigate how noise will distort slidograms and scalograms of
periodic functions. For this we have used our first artificial time-series
with added white-Gaussian noise with a mean of zero and a standard
deviation $\sigma_{noise}$ of 0.1. The power spectrum of these noisy
time-series (Fig.~\ref{test4wwz}, middle panel) show only
two test functions -- $F_1$ and $F_4$, while the detection of other
periodic signals is statistically insignificant. However, all seven test
functions are evident in both the slidogram and scalogram. For example, a
difference of the modulating frequencies $\nu_m$ of sinusoids $F_6$ and
$F_7$ can again be detected, although the latter look more
fragmentary, with a "ragged'' shape.

In conclusion we would like to point out some important features of the
test function $F_5$ in the scalogram in the presence of noise.
This function can be definitely detected, but the frequencies of its
spots changes considerably, as well as their brightness.
Occasionally some of them have completely disappeared.
On the whole, the shorter the life of the function the bigger
the variation in brightness and frequency.

\bsp
\label{lastpage}
\end{document}